\documentclass[journal]{IEEEtran}
\IEEEoverridecommandlockouts
\usepackage{cite}
\usepackage{tabularx,booktabs}
\usepackage{multirow}
\usepackage{amsmath,amssymb,amsfonts}
\usepackage{algorithmic}
\usepackage{graphicx}
\usepackage{textcomp}
\usepackage{xcolor}
\usepackage{mathtools}
\newcolumntype{Y}{>{\centering\arraybackslash}X}
\def\BibTeX{{\rm B\kern-.05em{\sc i\kern-.025em b}\kern-.08em
    T\kern-.1667em\lower.7ex\hbox{E}\kern-.125emX}}

\begin{document}

\title{A Dynamic Phasor Framework for Analysis of Subsynchronous Oscillations in Multi-Machine Systems with IBRs and Large Loads
}

\author{Fiaz Hossain,~\IEEEmembership{Student Member,~IEEE},  Nilanjan Ray Chaudhuri,~\IEEEmembership{Senior Member,~IEEE}, Constantino M. Lagoa,~\IEEEmembership{Member,~IEEE}, Alok Sinha, Sai Gopal Vennelaganti,~\IEEEmembership{Member,~IEEE}, and Mohammed E. Nassar,~\IEEEmembership{Senior Member,~IEEE}
\thanks{F. Hossain, N. R. Chaudhuri, and  C. M. Lagoa are with The School of Electrical Engineering and Computer
Science, Penn State University, University Park, PA, USA e-mail: fbh5142@psu.edu, nuc88@psu.edu, cml18@psu.edu.}
\thanks{A. Sinha is with Department of Mechanical Engineering, Penn State University, University Park, PA, USA e-mail: axs22@psu.edu.}
\thanks{S. G. Vennelaganti and M. E. Nassar are with Tesla Inc., Palo Alto, CA, USA e-mail: svennelaganti@tesla.com, monassar@tesla.com.}
\thanks{Financial support from NSF Grant Award ECCS 2317272 and from Tesla Inc.  under agreement 304869 PSNDA-0003632 is gratefully acknowledged.}
}

\maketitle 

\begin{abstract}
Although the electromagnetic transient (EMT) framework can capture subsynchronous oscillations (SSOs), it faces scalability issues for large-scale systems. Thus motivated, we propose a generalized dynamic phasor (DP) framework to analyze SSOs in  multi-machine systems with inverter-based resources (IBRs) and large loads such as artificial intelligence data centers (AI DCs) under balanced and unbalanced conditions. The grid-following (GFL) and grid-forming (GFM) IBRs are modeled in their respective $dq$-frame DPs. In contrast, the detailed model of multi-mass turbine driven synchronous generators (SGs) along with dynamic transmission network models and loads are represented in $pnz$-frame DPs. 
The linearizability and time-invariance of the framework enable us to perform eigen decomposition, which is a powerful tool for root-cause analysis of SSO modes and the design of damping controllers. In addition, the DP modeling approach facilitates faster simulation of large-scale systems. The generalized framework is validated with EMTDC/PSCAD simulations using the IEEE first benchmark model for subsynchronous resonance and the modified IEEE 4-machine system. Several use cases are presented on the modified IEEE 68-bus system with two GFL IBRs to show the applicability of the framework. First, time- and frequency-domain analyses of the IBR-induced SSO mode are presented. Then, two solutions are proposed to damp the poorly damped SSO mode: (a) a decentralized controller is designed using particle swarm optimization, and (b) the control of one GFL IBR is replaced by GFM control. Finally, the impact of AI DC load on primary frequency response of the system and the multi-mass turbines of the SGs are studied. 
\end{abstract}

\begin{IEEEkeywords}
Dynamic phasor, EMT, IBR, large load, AI data center, GFL, GFM, SSO, damping, shaft stress, fatigue life
\end{IEEEkeywords}

\section{Introduction}
\IEEEPARstart{S}{ome} of the unprecedented challenges that the power grid faces today demand scalable time- and frequency-domain analysis platforms. Subsynchronous oscillations (SSOs) driven by grid-following inverter-based resources (GFL IBRs) are increasingly observed in power systems \cite{Real_SSO_event}. This has led to the need for a system-level analysis of such phenomena and the design of damping controls \cite{Ameli-25-TPWRS}. Significant attention is also being paid to the relatively newer breed of grid-forming (GFM) IBRs with the promise of solving SSO problems. More recently, the rapid expansion of Artificial intelligence (AI) data center (DC) loads is creating potential reliability concerns \cite{NERC_2025}.  For example, large and steep power ramps are observed in AI DC loads during transitions from the training phase to the rest period and back, affecting the primary frequency response of the grid. Another significant challenge arises from the synchronized training cycles of AI data center loads, which induce large-amplitude fluctuations that may adversely affect the fatigue life of turbine-generator shafts. Although such effects can be modeled using state-of-the-art commercial electromagnetic transient (EMT) simulation tools, these approaches face limitations, including poor scalability due to high computational demands and a lack of suitability for developing linear time-invariant models required for frequency-domain analysis.

The dynamic phasor (DP)-based modeling approach \cite{Verghese-91-DP} serves as a complementary tool that can take advantage of adaptive time step solvers and achieve faster time-domain simulations than conventional EMT platforms. This is enabled by two key properties of the DP framework -- (a) users can choose the number of DP coefficients to restrict the range of captured frequency modes, and (b) the formulation is time-invariant. This framework can simulate large disturbances, including balanced and unbalanced short-circuit faults. The linearizability and time-invariant nature of DP models enable eigenvalue analysis to identify SSO modes, support root-cause analysis, and aid in the design of controllers, including damping controllers for these modes.

In the existing literature, the application of the DP framework has been largely restricted to systems comprising either a single machine or a single IBR; see, for instance, \cite{Stankovic_assymetric_SMIB,DP_SSR,hossain2025dynamicphasorframeworkanalysis}, which incorporate unbalanced conditions in their simulations. Although \cite{Vega_Herrera} reports the development of a DP-based model for a system consisting of one synchronous generator (SG) and two IBRs, the work uses the concepts of dynamic phasors and space phasors interchangeably. Furthermore, the proposed modeling approach assumes balanced operating conditions and provides limited detail on component-level models. A DP-based model of an aircraft power system is presented in \cite{YangTao}, where multiple SGs are connected in parallel; however, unlike terrestrial grids, the system does not include transmission lines or IBRs. Reference \cite{Demiray-08-DP-39bus} demonstrates a $10$x simulation speedup in a detailed DP-based model of the IEEE $39$-bus system  compared to the detailed EMT model. However, the paper does not include any modeling details of the DP framework and does not mention the simulation step sizes. Although this paper shows simulation results following unbalanced faults, the system does not consider IBRs. As such, the scalability of the DP framework for modeling multi-machine systems with IBRs under unbalanced conditions has hardly been demonstrated. 

Our recent work \cite{HossainChaudhuriLagoa2026DPSSO} represents an initial effort to address this gap. We study a modified IEEE $2$-area benchmark system with $2$ GFL IBRs that demonstrates IBR-induced SSO phenomena. The DP model shows a $2$x speed up on average compared to the EMT model with an averaged IBR representation, when a certain case is simulated $50$ times for $5$ s. We present a robust decentralized $\mathcal{H}_{\infty}$ SSO damping controller design approach using the linearized DP model and demonstrate its effectiveness following an unbalanced fault. Finally, we show that DC loads can excite the IBR-induced SSO mode and its locational impact on the excitation can be quantified using the frequency-dependent gains of the DP-based linearized input-output transfer function models.

In another recent work by the authors \cite{hossain2026limitingimpactaidata}, we proposed a framework to assess the impact of fluctuating AI DC loads on the fatigue life of the shafts of thermal turbine-generators in the subsynchronous frequency range. This framework needs lumped multi-mass models of turbine-generators, which was then validated using the DP-based model in the modified IEEE $4$-machine system and the modified IEEE 68-bus system. However, details of the DP-based modeling framework was not reported in \cite{hossain2026limitingimpactaidata}, since it was not the focus of that paper. 

The main contributions of this paper are the following.\\
1) We present a generalized DP-based modeling framework for analyzing phenomena in the subsynchronous range with the following capabilities and attributes --\\
a. multi-mass turbine-generator modeling \\
b. GFL and GFM IBR modeling\\
c. representation of balanced and unbalanced conditions\\
d. capturing transmission line dynamics using $\pi$ models\\
e. time-invariance, linearizability, tunable fidelity, scalability \\
2) We demonstrate benchmarking of DP models with EMT models and show speed up in a multi-machine system.\\
3) We show that the time-invariance and linearizability of the framework can be leveraged to design a supplementary low-order decentralized controller for damping a poorly-damped SSO mode in the modified IEEE-68 bus system with two GFL IBRs. We demonstrate that the damping controller performs well following different tie-line outages as a result of \textit{unbalanced faults}. \\
4)  We also show time- and frequency-domain analysis to demonstrate that changing the converter controls of one of the IBRs from GFL to GFM mode removes the poorly-damped SSO mode.\\
5) Finally, in the presence of the GFM IBR, we consider multi-mass turbine-generator shaft models. We present extensive simulation results to demonstrate the capability of the DP framework in analyzing impacts of AI DC loads on primary frequency response and torsional oscillations of the shear stress of the shaft sections, which can be instrumental in analyzing turbine-generator shaft fatigue life.  

We note that no novelty is claimed in designing the SSO damping control or the fact that the GFM control helps remedy the SSO problem. The contribution comes from showcasing the capability of the DP-based framework towards capturing these phenomena for the first time. The nomenclature and the parameter values used in the paper are given in the \textit{Supplementary Document.}


\begin{figure*}
	\centering
	\includegraphics[width= 0.8\textwidth]{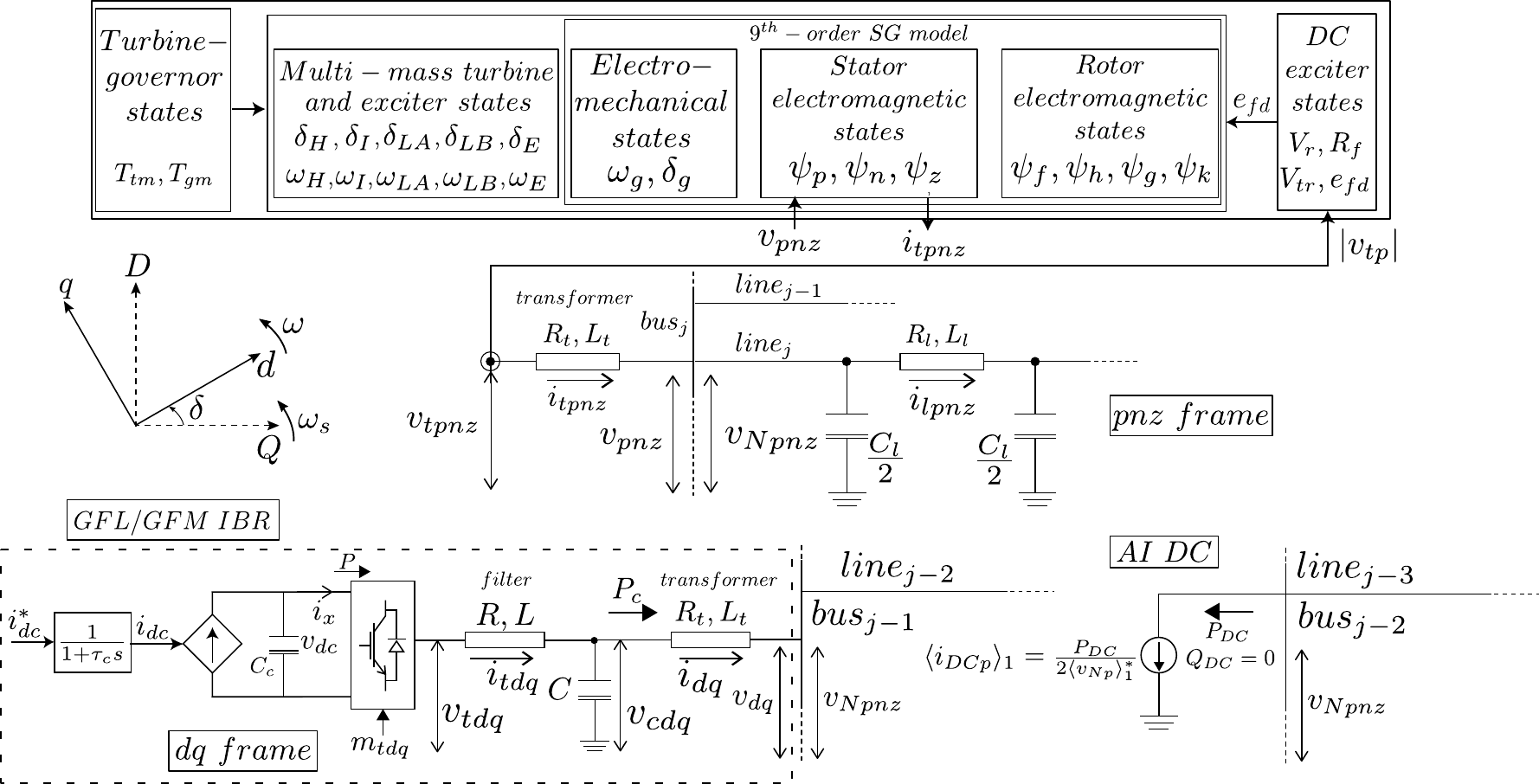}
	\caption{Proposed genralized DP-based modeling framework capable of analyzing balanced/unbalance conditions. 
    }
	\label{fig:DP_framework}
\end{figure*}

\section{DP-based Modeling Framework}\label{sec:DP_Framework}
In this section, we provide a concise overview of the proposed DP-based modeling framework. 

\subsection{Fundamentals of DP}
The concept of generalized averaging was originally introduced in \cite{Verghese-91-DP}. It states that a nearly periodic (and possibly complex-valued) signal $x(\tau)$ observed over the interval $\tau \in (t - T, t]$ can be represented as a Fourier series: $ x(\tau) = \sum_{k = -\infty}^{\infty} \langle x \rangle_k(t)\, e^{j k \omega_s \tau}$,
\normalsize
where $\omega_s = \frac{2\pi}{T}$, $k \in \mathbb{Z}$, and $X_k(t)$ denote time-varying complex Fourier coefficients obtained as the observation window of width $T$ moves along the signal.

Each coefficient $\langle x \rangle_k(t)$, referred to as the \textit{$k$th dynamic phasor (DP)}, is computed using the averaging operation \small
\[
\langle x \rangle_k(t) = \frac{1}{T} \int_{t-T}^{t} x(\tau)\, e^{-j k \omega_s \tau} \, d\tau 
.
\]
\normalsize

In practical applications of the DP framework, an approximation of $x(\tau)$ is formed by retaining only a subset $\mathcal{U}$ of dominant coefficients: \small
\[
x(\tau) \approx \sum_{k \in \mathcal{U}} \langle x \rangle_k (t)\, e^{j k \omega_s \tau}=\tilde{x}(\tau).
\] \normalsize
This truncation gives rise to a reduced-order model based on generalized averaging. For notational simplicity, the explicit time dependence of DPs will be omitted in the remainder of the discussion.

Some important properties of DPs are listed below:\\
(1) \small\[
\frac{d \langle x \rangle_k}{d t} 
= \left\langle \frac{d x}{d t} \right\rangle_k - j k \omega_s \langle x \rangle_k.
\]\normalsize 
(2)
If $x(\tau)$ is real-valued, then
\small\[
\langle x \rangle_k = \langle x \rangle_{-k}^*.
\]\normalsize
In the $pnz$ reference frame, this relation becomes
\small\[
\langle x_p \rangle_{-k} = \langle x_n \rangle_k^*.
\]\normalsize

The signal $x(\tau)$ may represent quantities in the $abc$ frame, an asynchronous $dq0$ frame, or a synchronous $DQ0$ frame (see the axis orientations in Fig.~\ref{fig:DP_framework}). By applying the transformation to a synchronous rotating reference frame, relationships between $DQ0$ and $pnz$ variables can be expressed in terms of dynamic phasors \cite{DP_SSR}.
\small
\begin{equation}\label{eqn:DQtopnz}
\resizebox{\columnwidth}{!}{$
    \begin{aligned}
        &\langle x_D\rangle_k=\frac{\langle x_p\rangle_{k+1}+\langle x_n\rangle_{k-1}}{\sqrt{2}};~
        \langle x_Q\rangle_k=-\frac{\langle x_p\rangle_{k+1}-\langle x_n\rangle_{k-1}}{\sqrt{2}j}\\
        &\langle x_p\rangle_{k+1}=\frac{\langle x_D\rangle_{k}-j\langle x_Q\rangle_{k}}{\sqrt{2}};~~
        \langle x_n\rangle_{k-1}=\frac{\langle x_D\rangle_{k}+j\langle x_Q\rangle_{k}}{\sqrt{2}};~\langle x_z \rangle_k = \langle x_0 \rangle_k 
    \end{aligned}
    $}
\end{equation}
\normalsize
The following equations describe the relationship between asynchronous $dq$ frame quantities and $pnz$ frame quantities
\vspace{-3pt}
\small
\begin{equation}\label{eqn:pnztodq}
\begin{aligned}
    &\langle x_d \rangle_k = \frac{1}{\sqrt{2}j}\left(e^{j\delta}\langle x_n \rangle_{k-1}-e^{-j\delta}\langle x_p \rangle_{k+1}\right)\\
    &\langle x_q \rangle_k = \frac{1}{\sqrt{2}}\left(e^{j\delta}\langle x_n \rangle_{k-1}+e^{-j\delta}\langle x_p \rangle_{k+1}\right) \\
    &\langle x_p \rangle_{k+1} = \frac{1}{\sqrt{2}}e^{j\delta}\left(\langle x_q \rangle_k-j\langle x_d \rangle_k\right)\\
    &\langle x_n \rangle_{k-1} = \frac{1}{\sqrt{2}}e^{-j\delta}\left(\langle x_q \rangle_k+j\langle x_d \rangle_k\right)
\end{aligned}
\end{equation}
\normalsize
where, $\delta$ is the angle between the $Q$-axis and the $d$-axis. 
Throughout the paper, the variables are defined as $\langle x_{dq}\rangle_k = [\langle x_{d}\rangle_k ~\langle x_{q}\rangle_k]^T$, $\langle x_{pnz}\rangle_k = [\langle x_{p}\rangle_k ~\langle x_{n}\rangle_k ~\langle x_{z}\rangle_k]^T$, and $[A]\langle x_{pnz}\rangle_k$ = $diag(a_p, a_n, a_z)\langle x_{pnz}\rangle_k$.

\subsection{Proposed generalized DP-based modeling framework capable of analyzing balanced/unbalanced conditions}
 
Our developed framework, illustrated in Fig.~\ref{fig:DP_framework}, incorporates the following key aspects:

   \noindent 1) \emph{IBR representation in the $dq$ reference frame:} 
    Since IBR controllers are commonly implemented using vector control in the $dq$ frame, it is natural to express the dominant DP coefficients in this same frame for both the averaged IBR circuit model and its control system, rather than transforming them into a different coordinate system. In practice, zero-sequence currents are typically suppressed in IBR operation. Therefore, we retain DP components corresponding to $k=0$ and $k=\pm2$, which capture the positive- and negative-sequence dynamics, respectively. For GFL IBRs, the $dq$ frame is defined by the phase-locked loop (PLL), whereas for GFM IBRs, it is governed by the frequency-droop mechanism \cite{hossain2025dynamicphasorframeworkanalysis}.

    \noindent 2)  \emph{Modeling of the network and SGs in the $pnz$ domain:} 
    Transmission systems, loads, and SGs are described in the $pnz$ domain. When analyzing phenomena near the nominal system frequency, such as SSOs, it is necessary to include DP components with $k=\pm1$. The mechanical dynamics of SGs, including multi-mass shaft systems, evolve slowly and can therefore be adequately represented using only the $k=0$ component. AI-based DC loads are modeled using positive-sequence quantities, represented by $k=\pm1$ DPs, since the case studies involving AI DCs assume balanced conditions.

    \noindent 3)  \emph{Coupling between IBRs and the external network:} 
    To interface IBRs with the rest of the system, DP quantities expressed in the $dq$ frame (such as terminal voltages and injected currents) are transformed into their $pnz$ counterparts using \eqref{eqn:pnztodq}. Conversely, variables represented in the $pnz$ frame can be mapped back into the $dq$ frame for use within IBR models.

\section{Component Modeling in DP Framework}\label{sec:DP_components}

\subsection{DP model of GFM IBR in $dq$-frame}\label{GFM IBR}
Under an ideal converter assumption, the GFM IBR structure is decomposed into two parts: (i) the droop mechanism and the outer voltage regulation loop, depicted in Fig. \ref{fig:GFC_control}(a), and (ii) the inner voltage loop, current regulation loop, and the output filter dynamics, illustrated in Figs \ref{fig:GFC_control}(b) and \ref{fig:GFC_control}(c), respectively. The DP representations of these subsystems are detailed in the following sections. 

\subsubsection{Outer voltage and droop controller (Fig.~\ref{fig:GFC_control}(a))} These components of the model produce the reference voltage and the asynchronous $dq$ frame angle relative to the synchronous $DQ$ frame. It can be represented within the DP framework using the following identities.
\small
\begin{equation}\label{eqn:outer}
\begin{aligned}
    &\langle \dot{v}_{cd}^* \rangle_0=K_{i,ac}\left(|\langle v_{cdq}^*\rangle_0|-|\langle v_{cdq}\rangle_0|\right)\\&\langle v_{cd}^* \rangle_0 = K_{p,ac}\left(|\langle v_{cdq}^*\rangle_0|-|\langle v_{cdq}\rangle_0|\right)\\&\langle P_c \rangle_0=\sum\limits_{k=0,\pm 2}\left(\langle v_{cd}\rangle_k\langle i_d\rangle_{-k}+\langle v_{cq}\rangle_k\langle i_q\rangle_{-k}\right)\\
    &\langle \dot{\tilde{P}}_c \rangle_0 = \frac{1}{\tau_p}\left(\langle P_c \rangle_0-\langle \tilde{P_c} \rangle_0\right);~
    \langle \dot{\delta}_c \rangle_0=d_{pc}\left(\langle P_c^* \rangle_0-\langle \tilde{P_c} \rangle_0\right)
\end{aligned}
\end{equation}
\normalsize

It is important to note that rapid fluctuations in $P_c(\tau)$ are suppressed by the low-pass filter, and the analysis retains only the $k = 0$ component in modeling the dynamics of $\langle \tilde{P}_c \rangle$. An analogous simplification is adopted for the dynamics of $\langle \delta_c \rangle$.

\begin{figure}
    \centering
    
    \begin{minipage}{0.45\textwidth}
        \centering
        \includegraphics[width=\textwidth]{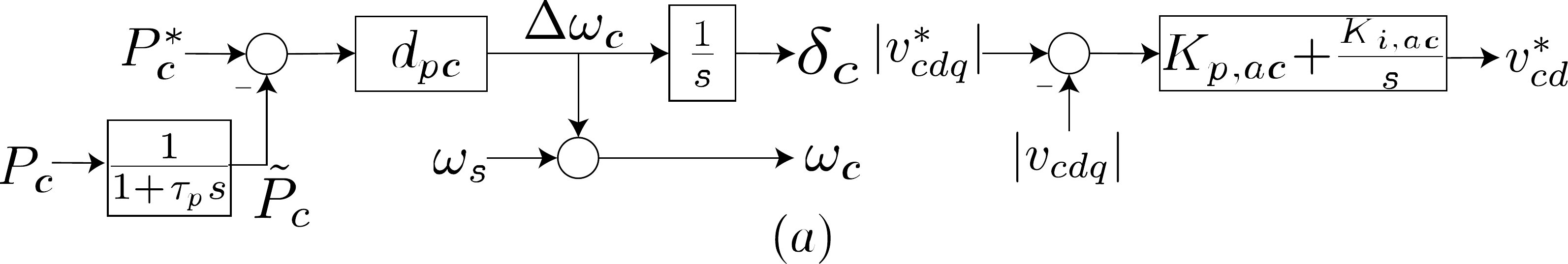}
        \vspace{-0.3cm}
    \end{minipage}
    
    
    \begin{minipage}{0.45\textwidth}
        \centering
        \includegraphics[width=\textwidth]{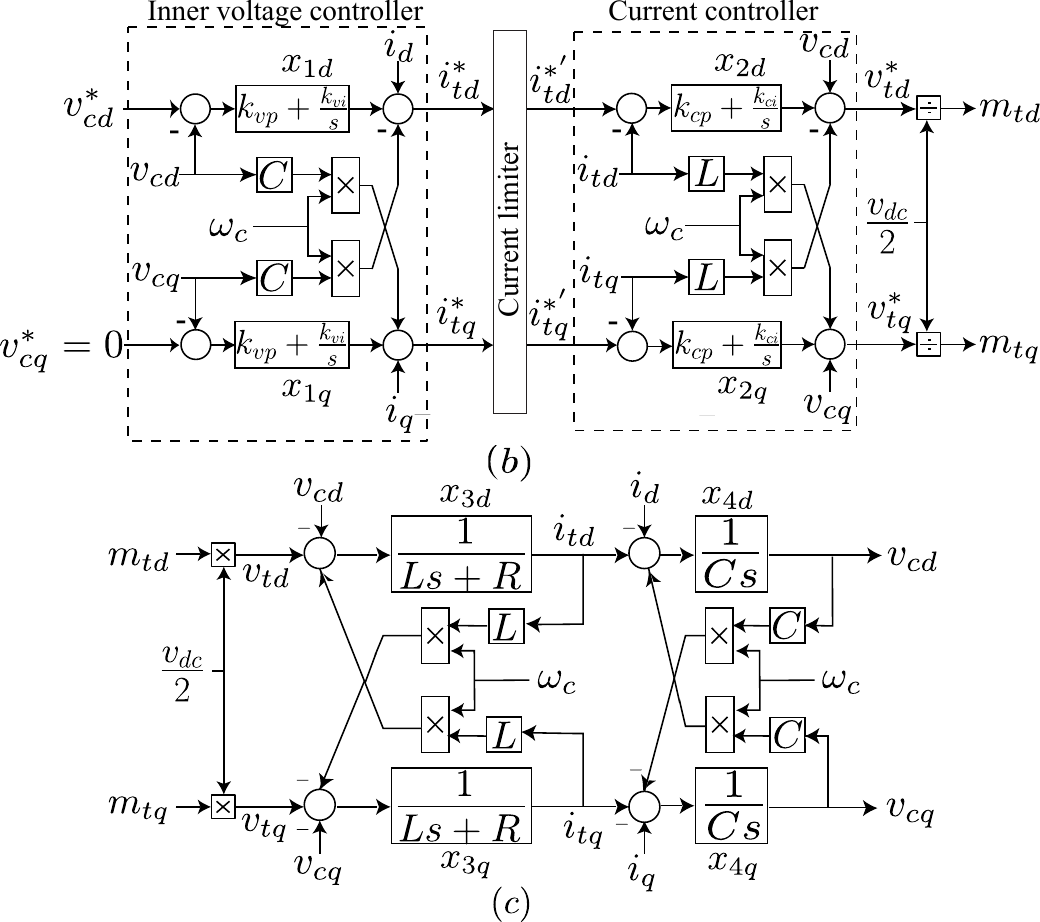}
        \vspace{-0.5cm}
    \end{minipage}
    
    \caption{Block diagram representation of the GFM IBR: (a) droop control and outer voltage controller, (b) inner voltage controller and current controller, and (c) output ac filter dynamics. Variables $x$ with subscripts denote the states of the corresponding blocks.}
    \label{fig:GFC_control}
     \vspace{-5pt}
\end{figure}

\subsubsection{Inner voltage controller (Fig.~\ref{fig:GFC_control}(b))} The state-space equations of this controller is expressed in the DP form as 
\small
\begin{equation}
\begin{aligned}    &\langle\dot{x}_{1dq}\rangle_k=B_{11}\langle v_{cdq}^*\rangle_k+B_{12}\langle v_{cdq}\rangle_k-jk\omega_s\langle x_{1dq}\rangle_k\\
    &\langle i_{tdq}^*\rangle_k=C_{11}\langle x_{1dq}\rangle_k+D_{11}\langle v_{cdq}^*\rangle_k+D_{12}\langle v_{cdq}\rangle_k+D_{13}\langle i_{dq}\rangle_k
\end{aligned}
\end{equation}
\normalsize
where, 
\small
\vspace{-3pt}
\begin{equation*}
\begin{aligned}
    &B_{11}=diag\left(1,1\right), ~B_{12}=-diag\left(1,1\right),~C_{11}=diag\left(k_{vi},k_{vi}\right)\\
    &D_{11}=diag\left(k_{vp},k_{vp}\right),D_{12}=\begin{bmatrix}-k_{vp} & -\omega_c C \\\omega_c C & -k_{vp} \end{bmatrix},D_{13} = diag(1,1).
\end{aligned}
\end{equation*}
\vspace{-3pt}
\normalsize



\begin{figure}
	\vspace{-5pt}
	\centering
\includegraphics[width= 0.45\textwidth]{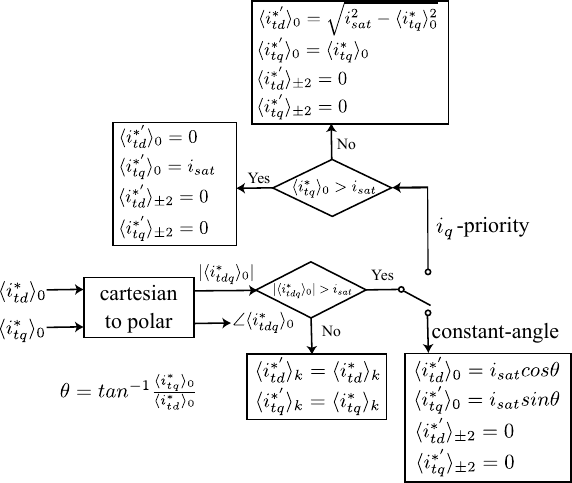}
	 \vspace{-5pt}
	\caption{Current limiting strategy in DP framework.}
\label{fig:curlim}
	 \vspace{-15pt}
\end{figure}

\subsubsection{Current limiter (Fig.~\ref{fig:GFC_control}(b))} 
A current-limiting mechanism is implemented between the voltage and current control loops to mitigate overcurrent during short-circuit conditions. Based on the operating state of the system, a constant-angle limiter or a $q$-component-current-priority ($i_q$-priority) limiter is engaged. In practical implementations, these approaches utilize a moving average filter to obtain the DC component of the angle $\theta$ derived from the $dq$-frame current signals. The current-limiting behavior as modeled within the DP framework is illustrated in Fig.~\ref{fig:curlim}. 
Here, $\theta$ is equal to $\tan^{-1}\left(\frac{\langle i_{tq}^*\rangle_0}{\langle i_{td}^*\rangle_0}\right)$, and $i_{sat}$ denotes $1.1$ times nominal current. 
\vspace{3pt}

\begin{figure}
	\centering
	\includegraphics[width= 0.47\textwidth]{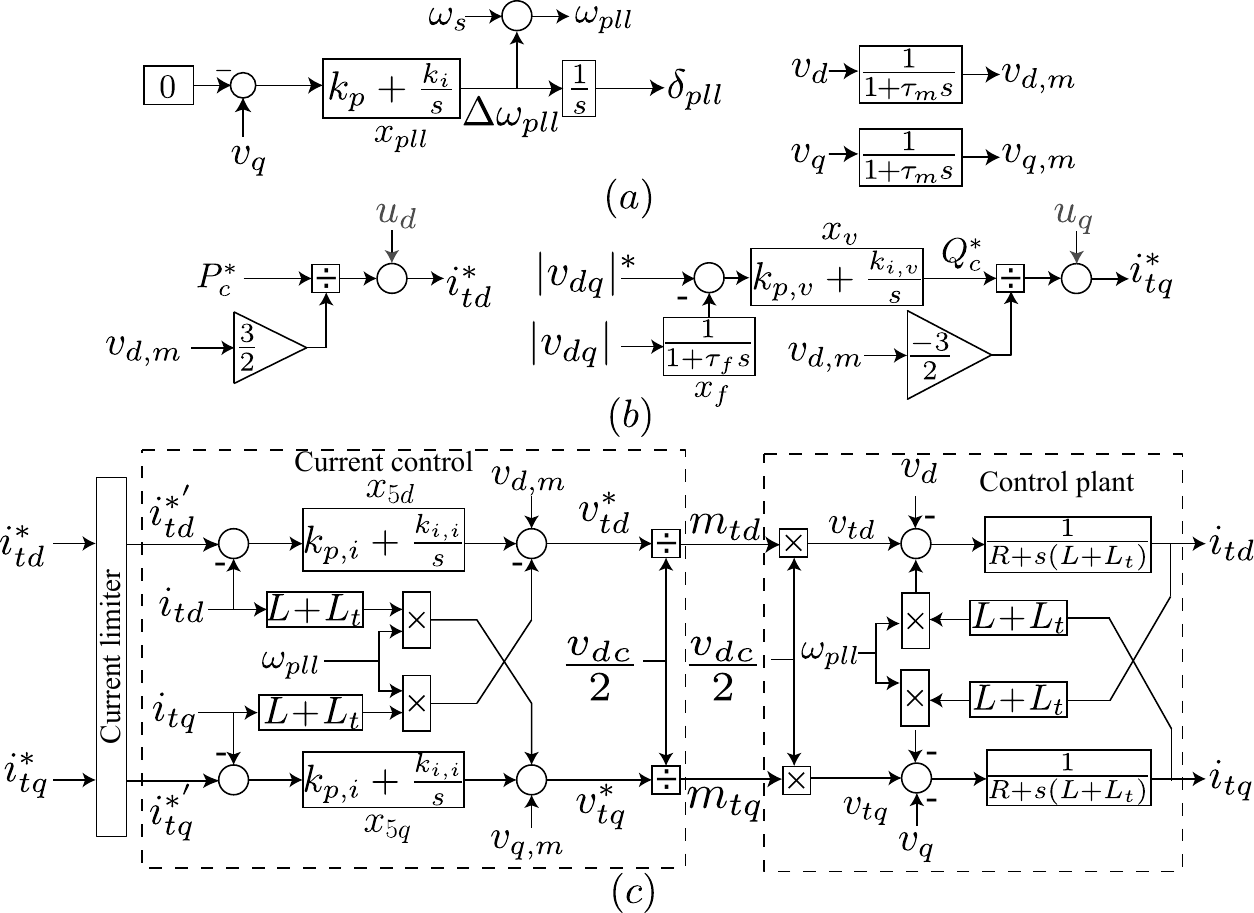}
	 \vspace{-10pt}
	\caption{Block diagram representation of the GFL IBR: (a) PLL, (b) outer control loops, and (c) inner control loops and control plant. Variables $x$ with subscripts denote the states of the corresponding blocks.}
	\label{fig:GFL IBR_control}
	 \vspace{-10pt}
\end{figure}
\subsubsection{Current controller (Fig.~\ref{fig:GFC_control}(b))} 
The following set of equations describes the current controller dynamics in DP form
\small
\vspace{-3pt}
\begin{equation}\label{eqn:curcontrol}
\begin{aligned}
&\langle\dot{x}_{2dq}\rangle_k=B_{21}\langle i_{tdq}^{*'}\rangle_k+B_{22}\langle i_{tdq}\rangle_k-jk\omega_s\langle x_{2dq}\rangle_k\\
    &\langle v_{tdq}^*\rangle_k=C_{21}\langle x_{2dq}\rangle_k+D_{21}\langle i_{tdq}^{*'}\rangle_k+D_{22}\langle i_{tdq}\rangle_k+D_{23}\langle v_{cdq}\rangle_k
\end{aligned}
\end{equation}
\normalsize
where,
\small
\vspace{-3pt}
\begin{equation*}
\begin{aligned}
    &B_{21}=diag\left(1,1\right), ~B_{22}=-diag\left(1,1\right),~C_{21}=diag\left(k_{ci},k_{ci}\right)\\
    &D_{21}=diag\left(k_{cp},k_{cp}\right),D_{22}=\begin{bmatrix}-k_{cp} & -\omega_c L \\\omega_c L & -k_{cp} \end{bmatrix},D_{23} = diag(1,1).
\end{aligned}
\end{equation*}
\normalsize

\subsubsection{Output filter dynamics (Fig.~\ref{fig:GFC_control}(c))} The RLC filter in GFM is modeled in DP form using the following equation
\small
\vspace{-3pt}
\begin{equation}
\begin{aligned}
   & \langle \dot{x}_{3dq}\rangle_k = A_{31}\langle x_{3dq}\rangle_k+B_{31}\langle v_{tdq}\rangle_k+B_{32}\langle v_{cdq}\rangle_k-jk\omega_s\langle x_{3dq}\rangle_k\\
   & \langle i_{tdq}\rangle_k=C_{31}\langle x_{3dq}\rangle_k\\
    &\langle \dot{x}_{4dq}\rangle_k = A_{41}\langle x_{4dq}\rangle_k+B_{41}\langle i_{tdq}\rangle_k+B_{42}\langle i_{dq}\rangle_k-jk\omega_s\langle x_{4dq}\rangle_k\\
   & \langle v_{cdq}\rangle_k=C_{41}\langle x_{4dq}\rangle_k
\end{aligned}
\end{equation}
\normalsize
where,
\small
\begin{equation*}
\begin{aligned}
    &A_{31}=\begin{bmatrix}-\frac{R}{L} & \omega_c \\-\omega_c & -\frac{R}{L} \end{bmatrix},~A_{41}=\begin{bmatrix}0 & \omega_c \\-\omega_c & 0 \end{bmatrix}\\&B_{31}=B_{41}=diag\left(1,1\right),~B_{32}=B_{42}=-diag\left(1,1\right)\\ &C_{31}=diag\left(\frac{1}{L},\frac{1}{L}\right),
    ~C_{41}=diag\left(\frac{1}{C},\frac{1}{C}\right).
\end{aligned}
\end{equation*}
\normalsize


\subsection{DP model of GFL IBR in $dq$-frame}
Under the assumption of an ideal converter, as illustrated in Fig.~\ref{fig:GFL IBR_control}, the GFL IBR can be represented by several subsystems: the PLL, outer control loops for active power and voltage regulation, an inner current control loop, and the plant. The plant comprises the series $R$--$L$ filter (with the shunt capacitor neglected) along with the transformer impedances shown in Fig.~\ref{fig:DP_framework}, where $i_{tdq} = i_{dq}$. 
The PLL dynamics depicted in Fig.~\ref{fig:GFL IBR_control}(a) are given below and are used to obtain the angle $\delta_{pll}$ between the $Q$- and $d$-axes illustrated in Fig.~\ref{fig:DP_framework}.

\small
\begin{equation}
    \begin{aligned}
        &\langle \dot{x}_{pll} \rangle_0 = k_{i}\langle v_q \rangle_0\\
        &\langle \dot{\delta}_{pll} \rangle_0 =  \Delta \omega_{pll}= k_{p}\langle v_q \rangle_0+\langle x_{pll} \rangle_0 
    \end{aligned}
\end{equation}

\normalsize

The dynamics of the outer control loops shown in Fig.~\ref{fig:GFL IBR_control}(b) are captured in the following DP models.
\small
\begin{equation}
    \begin{aligned}
        &\langle \dot{v}_{d,m} \rangle_k = \frac{1}{\tau_{m}}\left(\langle v_d \rangle_k-\langle v_{d,m} \rangle_k\right)-jk\omega_s\langle v_{d,m}\rangle_k\\
        &\langle \dot{v}_{q,m} \rangle_k = \frac{1}{\tau_{m}}\left(\langle v_q \rangle_k-\langle v_{q,m} \rangle_k\right)-jk\omega_s\langle v_{q,m}\rangle_k \\
        &\langle i_{td}^* \rangle_0 = \frac{2}{3}\frac{\langle P_c^* \rangle_0}{\langle v_{d,m} \rangle_0}+\langle u_d \rangle_0,~\langle i_{tq}^* \rangle_0 = -\frac{2}{3}\frac{\langle Q_c^* \rangle_0}{\langle v_{d,m} \rangle_0}+\langle u_q \rangle_0\\
        &\langle \dot{x}_f \rangle_0 = \frac{1}{\tau_f}\left( |\langle v_{dq}\rangle_0| -\langle x_f \rangle_0\right),~\langle \dot{x}_v \rangle_0 = k_{i,v}\left( |\langle v_{dq}^*\rangle_0| -\langle x_f \rangle_0\right)\\
        &\langle Q_c^* \rangle_0 = \langle x_v \rangle_0+k_{p,v}\left(|\langle v_{dq}^*\rangle_0|-\langle x_f \rangle_0\right) \\        
    \end{aligned}
\end{equation}
\normalsize

A constant-angle current limiting strategy is employed to restrict fault currents, as discussed in Subsection~\ref{GFM IBR}. The representation of the inner current control loops, together with the series $R$--$L$ filter interfacing the converter to the transformer (see Fig.~\ref{fig:GFL IBR_control}(c)), is illustrated below.
\small
\begin{equation}
\resizebox{\columnwidth}{!}{$
    \begin{aligned}
        &\langle \dot{x} _{5d} \rangle_k = k_{i,i}\left(\langle i_{td}^{*'} \rangle_k-\langle i_{td} \rangle_k\right)-jk\omega_s \langle x_{5d} \rangle_k\\
        &\langle \dot{x}_{5q} \rangle_k = k_{i,i}\left(\langle i_{tq}^{*'} \rangle_k-\langle i_{tq} \rangle_k\right)-jk\omega_s \langle x_{5q} \rangle_k \\
        &\langle v_{td}^* \rangle_k = k_{p,i}\left(\langle i_{td}^{*'} \rangle_k-\langle i_{td} \rangle_k\right)+\langle x_{5d} \rangle_k+\langle v_{d,m} \rangle_k-\omega_{pll}\left(L+L_t\right)\langle i_{tq} \rangle_k\\
        &\langle v_{tq}^{*} \rangle_k = k_{p,i}\left(\langle i_{tq}^{*'} \rangle_k-\langle i_{tq} \rangle_k\right)+\langle x_{5q} \rangle_k+\langle v_{q,m} \rangle_k+\omega_{pll} \left(L+L_t\right)\langle i_{td} \rangle_k\\
        &\langle \dot{i}_{td} \rangle_k = \frac{1}{L+L_t}\Big(\langle v_{td} \rangle_k-\langle v_{d} \rangle_k+\omega_{pll} \left(L+L_t\right)\langle i_{tq} \rangle_k\\
        &~~~~~~~~~ -\left(R+R_t\right)\langle i_{td} \rangle_k\Big)-jk\omega_s\langle i_{td} \rangle_k\\
        &\langle \dot{i}_{tq} \rangle_k = \frac{1}{L+L_t}\Big(\langle v_{tq} \rangle_k-\langle v_{q} \rangle_k-\omega_{pll}\left(L+L_t\right)\langle i_{td} \rangle_k\\
        &~~~~~~~~~ -\left(R+R_t\right)\langle i_{tq} \rangle_k\Big)-jk\omega_s\langle i_{tq} \rangle_k
    \end{aligned}
    $}
    \vspace{-10 pt}
\end{equation}
\normalsize

\subsection{DP model of SG in $pnz$ frame}
The equations of the complete rotor-mass system including high pressure (HP), intermediate pressure (IP), low pressure A (LPA), low pressure B (LPB) turbines, generator, and exciter (indexed as $r=6,5,\cdots,2,1$) are summarized below.
Let us assume that there are $N$ shaft sections connecting the $N$+$1$ masses. 
The state-space representation of the $r$th mass  can be expressed using the following set of equations.
\small
\begin{equation}\label{eq:Lumped}
\resizebox{\columnwidth}{!}{$
\begin{aligned}
    & \langle\dot{\delta}_r\rangle_0 =  \omega_s\left(\langle\omega_r\rangle_0 - 1\right) \\
    & \langle\dot{\omega}_r\rangle_0  =
    \frac{1}{2H_r}
    \left\{
    \begin{aligned}
        & \langle T_{mr}\rangle_0 - \langle T_e \rangle_0 - K_{r-1}\left( \langle\delta_r\rangle_0 - \langle\delta_{r-1}\rangle_0 \right)\\
        &- K_{r}\left( \langle\delta_r\rangle_0 - \langle\delta_{r+1}\rangle_0 \right)
        - D_{r-1}\left( \langle\omega_r\rangle_0 - \langle\omega_{r-1}\rangle_0 \right)\\
        &- D_{r}\left( \langle\omega_r\rangle_0 - \langle\omega_{r+1}\rangle_0 \right) 
        - \bar{D}_r \langle\omega_r\rangle_0 
    \end{aligned}
    \right\}
\end{aligned}
$}
\end{equation}
\normalsize

Note that $\langle T_{mr}\rangle_0$ is not present in generator and exciter mass, and $\langle T_{e}\rangle_0$ is only applicable for generator mass. The maximum tensile stress, $\sigma_{r}$ in the $r$th shaft section can be calculated using 
\begin{equation}\label{eq:Stress}
    \langle \sigma_{r}\rangle_0 = \frac{GR_r}{l_r}(\langle \theta_{r+1}\rangle_0-\langle\theta_{r}\rangle_0)
\end{equation}
\noindent where, $\langle\theta_r\rangle_0$ (mech. rad) is equal to $\frac{2}{p_f}\langle\delta_r\rangle_0$ and $p_f$ is the number of poles of the machine.

SG governor dynamics are modeled as follows. 

\small
\begin{equation}
\begin{aligned}
    &\langle u_g \rangle_0 = \frac{\omega_s}{2\pi R_{sp}}\left(1-\langle \omega_2 \rangle_0\right),~\langle \dot{T}_{gm} \rangle_0 = \frac{1}{\tau_g}\left(\langle u_g \rangle_0-\langle T_{gm} \rangle_0\right)\\
    &\langle \dot{T}_{tm} \rangle_0 = \frac{1}{\tau_t}\left(\langle T_{gm} \rangle_0-\langle T_{tm} \rangle_0\right),~\langle T_{mr} \rangle_0 =F_r\left( \langle T_{m} \rangle_0+\langle T_{tm} \rangle_0\right)\\
\end{aligned}
\end{equation}
\normalsize

The stator transients are modeled using the following equation in pnz domain.
\small
\begin{equation}
    \langle \dot{\psi}_{pnz} \rangle_k = -\langle v_{pnz} \rangle_k-[R_s] \langle i_{tpnz} \rangle_k-jk\omega_s\langle \psi_{pnz} \rangle_k
\end{equation}
\normalsize

In addition to the field winding, one damper along the $d$-axis and two dampers along the $q$-axis are modeled, and $k=0,\pm 2$ DPs are considered to model the rotor circuit dynamics.  
\small
\begin{equation}
    \langle \dot{\psi}_{r} \rangle_k = -\langle v_{r} \rangle_k-[R_r] \langle i_{r} \rangle_k-jk\omega_s\langle \psi_{r} \rangle_k
\end{equation}
\normalsize 

Flux linkages between stator and rotor can be expressed using the following equations.
\small
\begin{equation}\label{eq:FluxLink}
\resizebox{\columnwidth}{!}{$
    \begin{bmatrix}
        \langle\psi_{pnz}^{'}\rangle_1\\
        \langle \psi_{pnz}^{'}\rangle_{-1}\\
        \langle\psi_{r}^{'}\rangle_0\\
        \langle\psi_{r}^{'}\rangle_2\\
        \langle\psi_{r}^{'}\rangle_{-2}
    \end{bmatrix}
    =
    \begin{bmatrix}
        \langle L_{ss}^{'}\rangle_{0} & \langle L_{ss}^{'}\rangle_{2} & \langle L_{sr}^{'}\rangle_{1} & \langle L_{sr}^{'}\rangle_{-1} & 0\\
        \langle L_{ss}^{'}\rangle_{-2} & \langle L_{ss}^{'}\rangle_{0} & \langle L_{sr}^{'}\rangle_{-1} & 0 & \langle L_{sr}^{'}\rangle_{1}\\
        \langle L_{rs}^{'}\rangle_{-1} & \langle L_{rs}^{'}\rangle_{1} & \langle L_{rr}\rangle_{0} & 0 & 0\\
        \langle L_{rs}^{'}\rangle_{1} & 0 & 0 & \langle L_{rr}\rangle_{0} & 0\\
        0 & \langle L_{rs}^{'}\rangle_{-1} & 0 & 0 & \langle L_{rr}\rangle_{0}
    \end{bmatrix}
    \begin{bmatrix}
        \langle i_{tpnz}^{'}\rangle_1\\
        \langle i_{tpnz}^{'}\rangle_{-1}\\
        \langle i_{r}^{'}\rangle_0\\
        \langle i_{r}^{'}\rangle_2\\
        \langle i_{r}^{'}\rangle_{-2}
    \end{bmatrix}
    $}
\end{equation}
\normalsize
where,
\small
\begin{equation*}
\resizebox{\columnwidth}{!}{$
\begin{aligned}
    &\langle L_{ss}\rangle_0=
    \begin{bmatrix}
        L_{aa0} & L_{ab0} & L_{ab0} \\
        L_{ab0} & L_{aa0} & L_{ab0} \\
        L_{ab0} & L_{ab0} & L_{aa0}
    \end{bmatrix},~
    \langle L_{rr}\rangle_0=
    \begin{bmatrix}
        L_{f} & L_{fh} & 0 & 0 \\
        L_{fh} & L_{h} & 0 & 0 \\
        0 & 0 & L_{g} & L_{gk} \\
        0 & 0 & L_{gk} & L_{k}
    \end{bmatrix}\\
    &\langle L_{ss}\rangle_2= \frac{L_{aa2}}{2}
    \begin{bmatrix}
        1 & \alpha^* & \alpha \\
        \alpha^* & \alpha & 1 \\
        \alpha & 1 & \alpha^*
    \end{bmatrix},~
    \langle L_{sr}\rangle_1= \frac{1}{2}
    \begin{bmatrix}
        M_{af} & M_{ah} & -jM_{ag} & -jM_{ak} \\
        M_{af}\alpha^* & M_{ah}\alpha^* & -jM_{ag}\alpha^* & -jM_{ak}\alpha^* \\
        M_{af}\alpha & M_{ah}\alpha & -jM_{ag}\alpha & -jM_{ak}\alpha
    \end{bmatrix}\\
    &\langle L_{ss}^{'} \rangle_k = T^{-1}\langle L_{ss} \rangle_k T,~\langle L_{sr}^{'} \rangle_k = T^{-1}\langle L_{sr} \rangle_k,~\langle L_{rs}^{'} \rangle_k = \langle L_{rs} \rangle_k T,~\langle L_{sr} \rangle_k = \langle L_{rs} \rangle_k^{T},\\
    &~T = \frac{1}{\sqrt{3}}\begin{bmatrix}
        1 & 1 & 1\\
        \alpha^2 & \alpha & 1\\
        \alpha & \alpha^2 & 1
    \end{bmatrix}, ~\alpha=e^{j\frac{2\pi}{3}}
\end{aligned}
$}
\end{equation*}
\normalsize

It is worth mentioning that the fluxes and currents of the stator and the rotor are transformed using $\langle f' \rangle_k = \langle f \rangle_k e^{-jk\delta_g}$ to make the coupling inductance matrix time-invariant. The electromagnetic torque can be expressed as follows.
\vspace{-5pt}
\small
\begin{equation}
    \langle T_e \rangle_k = \sum\limits_{l=-\infty}^{\infty}\langle \psi_D \rangle_{k-l}\langle i_{tQ} \rangle_l-\sum\limits_{l=-\infty}^{\infty}\langle \psi_Q \rangle_{k-l}\langle i_{tD} \rangle_l.
\end{equation}
\normalsize

SGs are equipped with IEEE DC1A exciters \cite{DC1A}, which are modeled considering only the $0$th DP. 
\small
\begin{equation}
\resizebox{\columnwidth}{!}{$
    \begin{aligned}
        &\langle \dot{R}_f \rangle_0 = \frac{1}{T_F}\left(\langle e_{fd} \rangle_0-\langle R_f \rangle_0\right),~\langle \dot{V}_{tr} \rangle_0 = \frac{1}{T_r}\left(|\langle v_{tp}\rangle_1|-\langle V_{tr} \rangle_0\right)\\
        &\langle \dot{V}_{r} \rangle_0 = \frac{1}{T_A}\Bigg\{\frac{K_AK_F}{T_F}\left(\langle R_f \rangle_0-\langle e_{fd} \rangle_0\right)+K_A \left(\langle v_{ref} \rangle_0-\langle V_{tr} \rangle_0\right)-\langle V_{r} \rangle_0\Bigg\}\\
        &\langle \dot{e}_{fd} \rangle_0 = -\frac{1}{T_E}\Big\{K_E \langle e_{fd} \rangle_0+A_{ex}e^{B_{ex}\langle e_{fd} \rangle_0}\langle e_{fd} \rangle_0-\langle V_{r} \rangle_0\Big\}\\
        &\langle v_{f} \rangle_0 = \frac{R_{f}}{L_{adu}}\langle e_{fd} \rangle_0 
    \end{aligned}
    $}
\end{equation}
\normalsize

Static exciters with power system stabilizer (PSS) can be modeled in DP using the following equations.
\vspace{-5pt}
\small
\begin{equation}
\vspace{-10pt}
\resizebox{\columnwidth}{!}{$
    \begin{aligned}
        & \langle \dot{V}_{tr} \rangle_0 = \frac{1}{T_r}\left(|\langle v_{tp}\rangle_1|-\langle V_{tr} \rangle_0\right),~ \langle\dot{x}_s \rangle_0 = \frac{1}{T_s}(\langle \omega_r \rangle_0-\langle x_s \rangle_0) \\
        & \langle \dot{x}_{wo} \rangle_0 = \frac{1}{T_{w}}(\langle x_s \rangle_0-\langle x_{wo}\rangle_0),~\langle \dot{x}_{comp1} \rangle_0 = \frac{1}{T_2}\left( \langle x_s \rangle_0 - \langle x_{wo} \rangle_0 -\langle x_{comp1} \rangle_0 \right) \\
        & \langle \dot{x}_{comp2} \rangle_0 = \frac{1}{T_4}\Bigg\{ \left(1-\frac{T_1}{T_2}\right) \langle x_{comp1} \rangle_0 + \frac{T_1}{T_2}\left( \langle x_s \rangle_0 - \langle x_{wo} \rangle_0\right) - \langle x_{comp2} \rangle_0 \Bigg\} \\
        & \langle e_{fd} \rangle_0 = K_A\Bigg[K_{PSS}\Bigg\{ \left(1-\frac{T_3}{T_4}\right) \langle x_{comp2} \rangle_0 + \frac{T_3}{T_4}\left(1-\frac{T_1}{T_2}\right) \langle x_{comp1} \rangle_0 \\
        &~~~~~+ \frac{T_1 T_3}{T_2 T_4}\left( \langle x_s \rangle_0 - \langle x_{wo} \rangle_0\right) \Bigg\}+ \langle v_{ref} \rangle_0-\langle V_{tr} \rangle_0\Bigg],~\langle v_{f} \rangle_0 = \frac{R_{fd}}{L_{adu}}\langle e_{fd} \rangle_0
    \end{aligned}
    $}
\end{equation}
\normalsize

\subsection{DP model of transmission network in $pnz$ frame}
The transmission network considers a lumped $\pi$-section model consisting of the following KCL and KVL algebraic equations.
\small
\begin{equation}
\resizebox{\columnwidth}{!}{$
\begin{aligned}
    &\langle i_{Npnz} \rangle_k = CCI \times [\langle i_{tpnz}\rangle_k^T  ~\langle i_{lpnz}\rangle_k^T]^T,~\langle v_{lpnz} \rangle_k = CCU \times \langle v_{Npnz}\rangle_k
\end{aligned}
$}
\end{equation}
\normalsize
Assuming the network has $n$ nodes, $l$ series $R-L$ branches excluding $m$ SG, $p$ GFM IBR and $q$ GFL IBR transformers, $\langle i_{Npnz} \rangle_k \in \mathbb{C}^{3n}$ is the vector of net injected currents in the nodes going towards shunt capacitances and any load that may be present, $\langle i_{lpnz}\rangle_k \in \mathbb{C}^{3l}$ and $\langle i_{tpnz}\rangle_k \in \mathbb{C}^{3(m+p+q)}$ are the vectors of currents flowing through each series $R-L$ branches and transformers, $\langle v_{Npnz}\rangle_k \in \mathbb{C}^{3n}$ is the node voltage vector, $\langle v_{lpnz}\rangle_k \in \mathbb{C}^{3l}$ are the voltages across $R-L$ branches, and $CCI \in \mathbb{R}^{3n \times (3(l+m+p+q))}$ and $CCU \in \mathbb{R}^{3l \times 3n}$ are the incidence matrix and nodal connectivity matrix, respectively. The transmission network is interfaced with SGs and GFL IBRs according to Fig. \ref{fig:DP_framework}. 
\small
\begin{equation}
\begin{aligned}
    &\langle \dot{i}_{lpnz} \rangle_k = [L_l]^{-1}\left(\langle v_{lpnz}\rangle_k-[R_l]\langle i_{lpnz} \rangle_k-jk\omega_s[L_l]\langle i_{lpnz} \rangle_k\right)\\
    &\langle \dot{v}_{Npnz} \rangle_k = [C_l]^{-1}\left(\langle i_{cpnz} \rangle_k-jk\omega_s[C_l]\langle v_{Npnz} \rangle_k\right)
\end{aligned}
\vspace{-10 pt}
\end{equation}
\normalsize
\begin{figure}
	\centering
	\includegraphics[width= 0.45\textwidth]{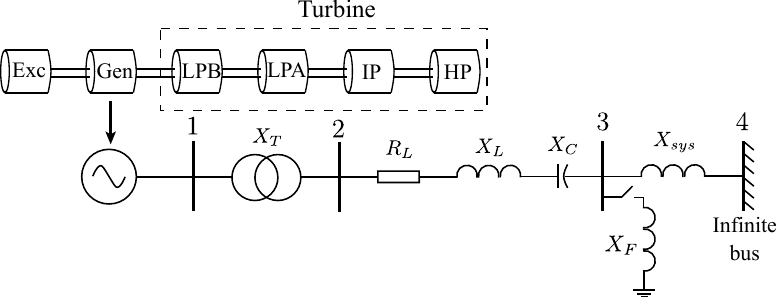}
	 \vspace{-7pt}
	\caption{IEEE First benchmark model for SSR \cite{IEEESSRBenchmark1977}.}
	\label{fig:IEEE FBM}
	\vspace{-10pt}
\end{figure}
\subsection{DP model of loads in $pnz$ frame}
At the load buses, constant impedance loads are modeled dynamically as parallel combinations of $R_L$, $L_L$, and $C_L$ elements.
\small
\begin{equation}
    \begin{aligned}
        &\langle \dot{i}_{LLpnz} \rangle_k = [L_L]^{-1}\left(\langle v_{Npnz}\rangle_k-jk\omega_s[L_L]\langle i_{LLpnz} \rangle_k\right)\\
        &\langle i_{LRpnz} \rangle_k = [R_L]^{-1}{\langle v_{Npnz}\rangle_k},~\langle i_{Lpnz} \rangle_k  = \langle i_{LLpnz} \rangle_k+\langle i_{LRpnz} \rangle_k
    \end{aligned}
\end{equation}
\normalsize

DC loads are represented as constant power demands with unity power factor  in the model. Furthermore, it is assumed that these loads do not contribute any negative- or zero-sequence current components.
\small
\begin{equation}\label{eq:DC_DP}
    \begin{aligned}
        \langle i_{DCp} \rangle_1 = \frac{\langle P_{DC}\rangle_0}{2\langle v_{Np}\rangle_1^{*}},~\langle i_{DCn} \rangle_1=0,~\langle i_{DCz} \rangle_1=0
    \end{aligned}
\end{equation}
\normalsize
\subsection{Fault and line outage modeling}
The resistive fault is represented through the bus fault impedance matrix formulated in the $pnz$ reference frame, which can be written as
\small
\begin{equation*}
\resizebox{\columnwidth}{!}{$
\begin{aligned}
    &R_{fabcg} = \begin{bmatrix} R_{fa}+R_g & R_g & R_g\\R_g & R_{fb}+R_g & R_g\\R_g & R_g & R_{fc}+R_g\end{bmatrix},~R_{fpnz} = T^{-1}R_{fabcg}T.\\
\end{aligned}
$}
\end{equation*}
\normalsize
Observe that $R_{fa}$ and $R_{g}$ denote, respectively, the phase-$a$ to neutral resistance (with analogous definitions for phases $b$ and $c$) and the neutral-to-ground fault resistance. 

As an illustration, a phase-$a$ to ground fault can be represented by choosing
\[
R_{fa} = R_f, \quad R_{fb} \to \infty, \quad R_{fc} \to \infty, \quad R_g = 0,
\]
where, $R_f$ characterizes the fault resistance. In numerical implementations, the infinite resistances $R_{fb}$ and $R_{fc}$ are approximated by sufficiently large values.

The resulting fault current is incorporated into the model through the following set of equations.
\small
\begin{equation}
    \begin{aligned}
        &\langle i_{fault} \rangle_k = R_{fpnz}^{-1}\langle v_{Npnz} \rangle_k \\
        &\langle i_{cpnz} \rangle_k = \langle i_{Npnz} \rangle_k-\langle i_{Lpnz} \rangle_k-\langle i_{DCpnz} \rangle_k-\langle i_{fault} \rangle_k
    \end{aligned}
\end{equation}
\normalsize

Line outage is simulated by modeling circuit breaker opening in both ends of the line. The opening of a circuit breaker is simulated by increasing the series resistance from a small value (e.g., $10^{-10}\Omega$) to a high value (e.g., $10^{10}\Omega$) within $3$ to $5$ cycles. 
\vspace{-11pt}
\section{Simulation Results and Discussions}
In this section, we present a benchmarking exercise of three DP models with their EMT counterparts to assess different aspects. After gaining confidence in the framework, we consider the SSO damping problem and the evaluation of impact of AI DC loads on turbine-generator shaft stress as use cases in the modified IEEE 68-bus system. DP models were built in MATLAB/Simulink \cite{matlab} and run using a variable timestep solver \textit{ode23tb} on a 64-bit Windows workstation with Intel® Xeon® W-2123 processor (3.60 GHz) and 48 GB RAM. 
\vspace{-10pt}


\begin{table}
\centering
\caption{Torsional and network modes for IEEE FBM}\label{tab:linearization}
\begin{tabular}{|c|c|c|c|}
\hline
\multicolumn{2}{|c|}{Torsional modes} & \multicolumn{2}{c|}{Network modes} \\
\hline
DP model & \cite{DP_SSR} & DP model & \cite{DP_SSR} \\
\hline
0.02$\pm$j99.5   & 0.02$\pm$j99.5   & -3.70$\pm$j155.65 & -3.70$\pm$j155.64 \\
0.08$\pm$j127.1  & 0.08$\pm$j127.1  & -4.69$\pm$j155.38 & -4.71$\pm$j154.56 \\
0.59$\pm$j159.8  & 0.59$\pm$j159.8  & -53.55$\pm$j230.85 & -53.58$\pm$j230.81 \\
0.004$\pm$j202.8 & 0.005$\pm$j202.8 & -53.55$\pm$j523.15 & -53.58$\pm$j523.17 \\
$\pm$j298.2       & $\pm$j298.2       & -4.69$\pm$j598.83 & -4.69$\pm$j598.82 \\
                  &                   & -3.02$\pm$j598.62 & -3.09$\pm$j599.39 \\
\hline
\end{tabular}
\end{table}

\begin{figure}[!t]
	\vspace{-10pt}
	\centering
	\includegraphics[trim = {1.3cm 6.4cm 1.3cm 6.5cm}, clip,width= 0.4\textwidth]{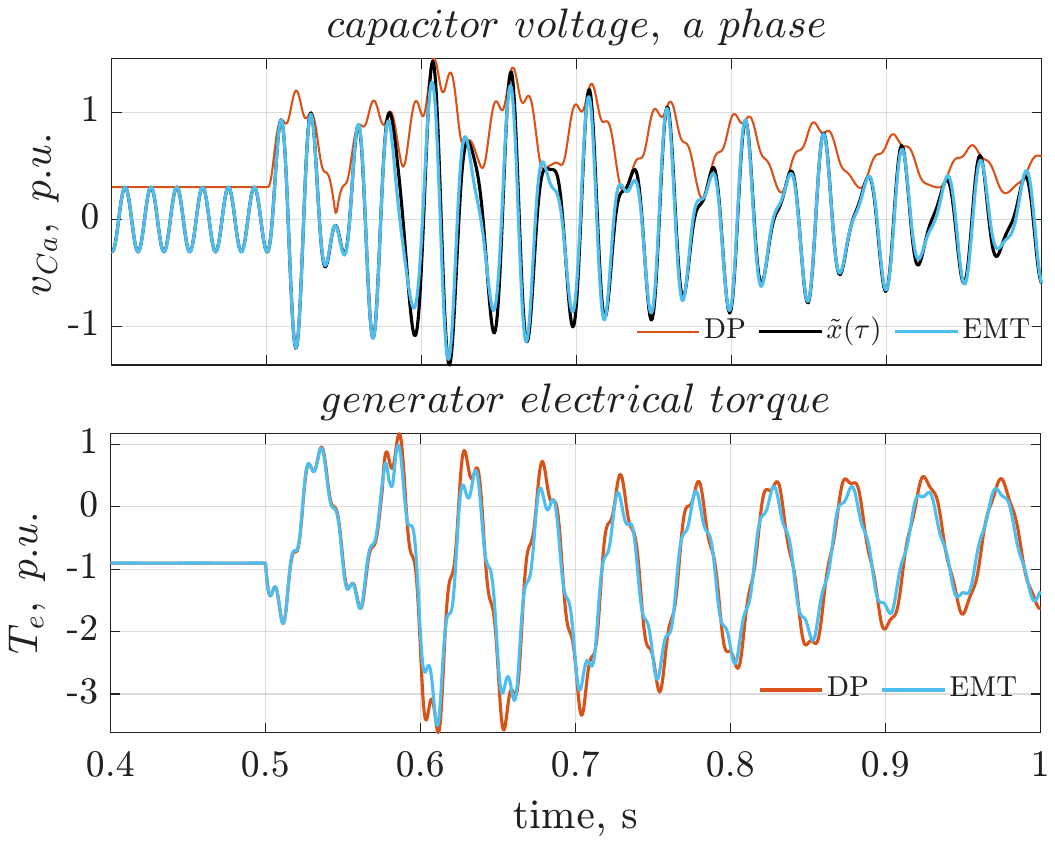}
	 \vspace{-15pt}
	\caption{Comparison of responses from DP model and EMT model of IEEE FBM. }
    \label{fig:dp_emt_validation}
\end{figure}

\begin{figure}[!t]
	\vspace{-5pt}
	\centering
	\includegraphics[width= 0.35\textwidth]{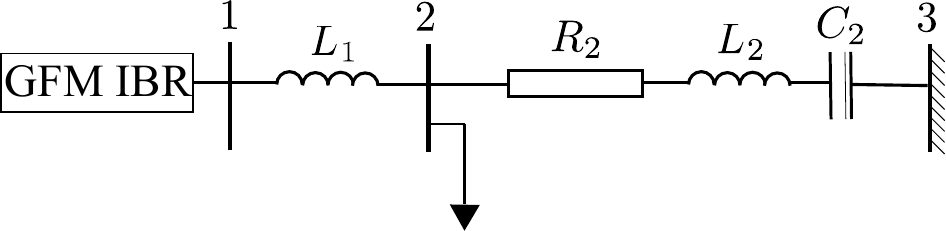}
	 \vspace{-10pt}
	\caption{GFM IBR connected with series compensated line.}
	\label{fig:GFC_SLD}
	\vspace{-10pt}
\end{figure}

\begin{figure}
    \centering
    
    \begin{minipage}{0.4\textwidth}
        \centering
        \includegraphics[trim={1.5cm 6.2cm 1.6cm 6.8cm},clip,width=\textwidth]{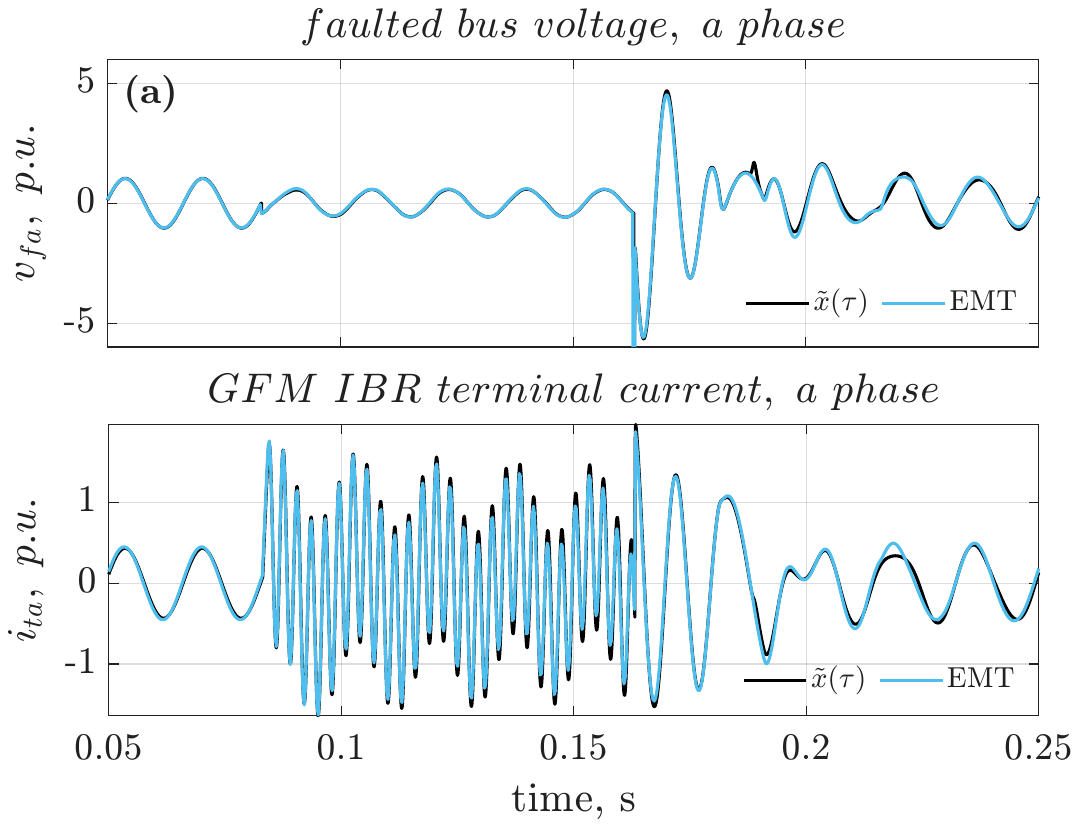}
        \vspace{-0.8cm}
    \end{minipage}
    
    
    \begin{minipage}{0.4\textwidth}
        \centering
        \includegraphics[trim={1.5cm 6.2cm 1.6cm 6.5cm},clip,width=\textwidth]{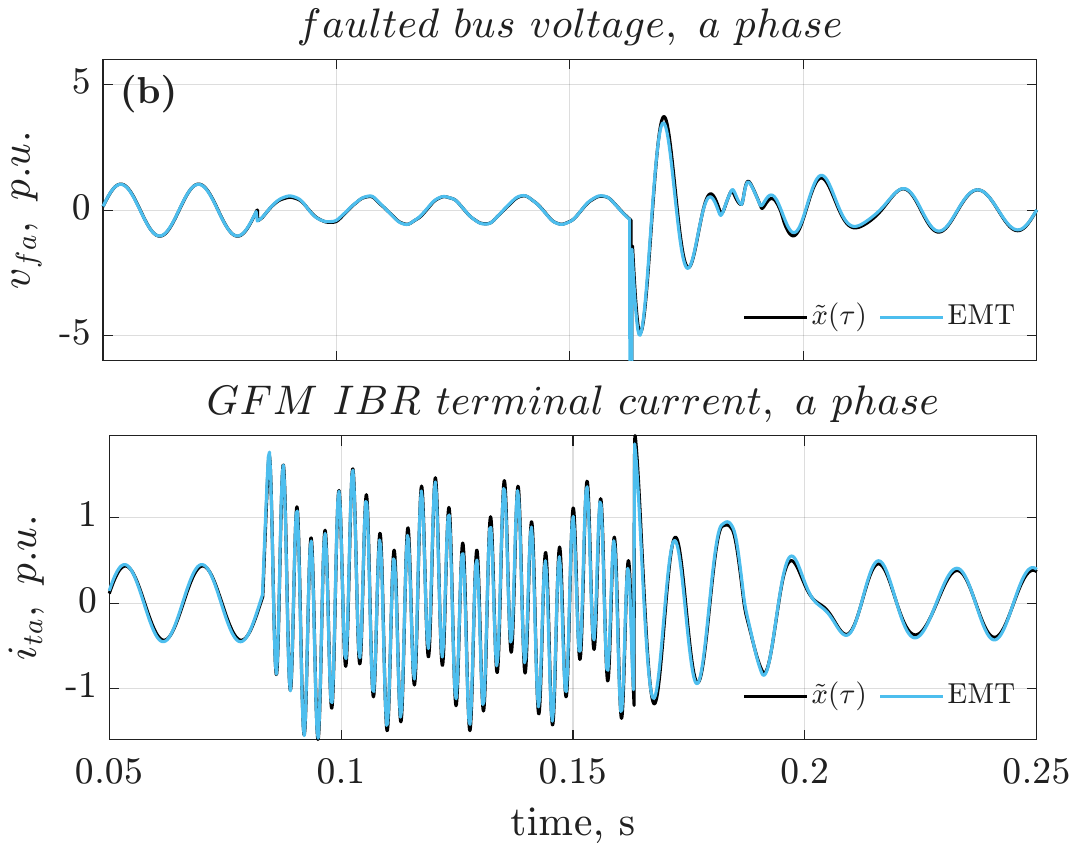}
        \vspace{-0.8cm}
    \end{minipage}
    
    \caption{Comparison of responses from DP and EMT model following a $5$ cycles self-clearing L-L fault near bus $2$ in Fig. \ref{fig:GFC_SLD} while (a) constant angle and (b) $i_q$- priority current limiter is active, respectively.}
    \label{fig:GFM_DP_EMT}
    \vspace{-10pt}
\end{figure}

\begin{figure}
	\centering
	\includegraphics[width= 0.45\textwidth]{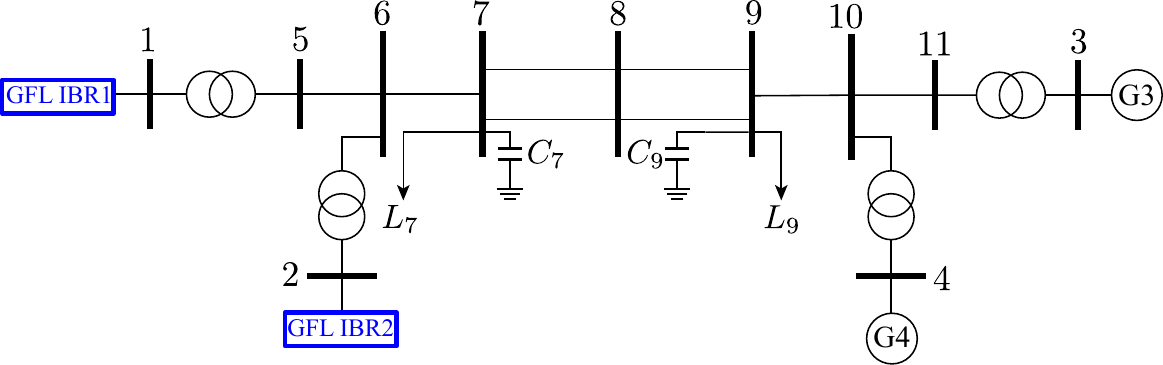}
	 \vspace{-5pt}
	\caption{Modified IEEE 4-machine system \cite{hossain2025dynamicphasorframeworkanalysis}.}
	\label{fig:IEEE 4-machine system}
	\vspace{-10pt}
\end{figure}

\subsection{Validation of the DP Framework}\label{sec:DP_Valid}
To gain confidence in the DP model, we perform three stages of validation with respect to EMT models.
\subsubsection{Validation of DP-based multi-mass SG model and torsional interaction}
To validate the modeling approach for the multi-mass turbine based SG using the DP framework, we have performed both frequency- and time-domain analyses in the IEEE First benchmark model (FBM) for SSR \cite{IEEESSRBenchmark1977} as shown in Fig. \ref{fig:IEEE FBM}. The eigenvalues of the linearized DP model are shown in Table \ref{tab:linearization} and compared to the eigenvalues presented in \cite{DP_SSR}. Furthermore, a three phase-to-ground (LLL-G) fault is applied near bus $3$ in both the DP and a publicly available EMT model of the system in Matlab/Simscape \cite{Young2026SSR}, which is run using a variable timestep solver \textit{ode23tb}. The simulation results are compared in Fig. \ref{fig:dp_emt_validation}, where a close match is observed between the responses. 
\begin{table}[!t]
\caption{SSO mode in different modeling frameworks}
\label{tab:SSO_comparison}
\centering
\resizebox{0.40\textwidth}{!}{
\begin{tabular}{|c|c|c|c|c|}
\hline
\textbf{PLL BW} & \multicolumn{2}{c|}{\textbf{15 Hz}} & \multicolumn{2}{c|}{\textbf{20 Hz}}\\ \hline
\textbf{Framework} & \textbf{EMT} & \textbf{DP} & \textbf{EMT} & \textbf{DP}\\ \hline
Approach & Prony & Linearization & Prony & Linearization\\ \hline
$f,\ \mathrm{Hz}$ & 5.238 & 5.156 & 6.383 & 6.357\\ \hline
$\zeta,\ \%$ & 15.3 & 16.3 & 1.2 & 0.2\\ \hline

\end{tabular}
}
\vspace{-12pt}
\end{table}
\begin{figure}[!t]
	\vspace{-5pt}
	\centering
	\includegraphics[trim = {1.2cm 6.5cm 1.6cm 6cm}, clip,width= 0.43\textwidth]{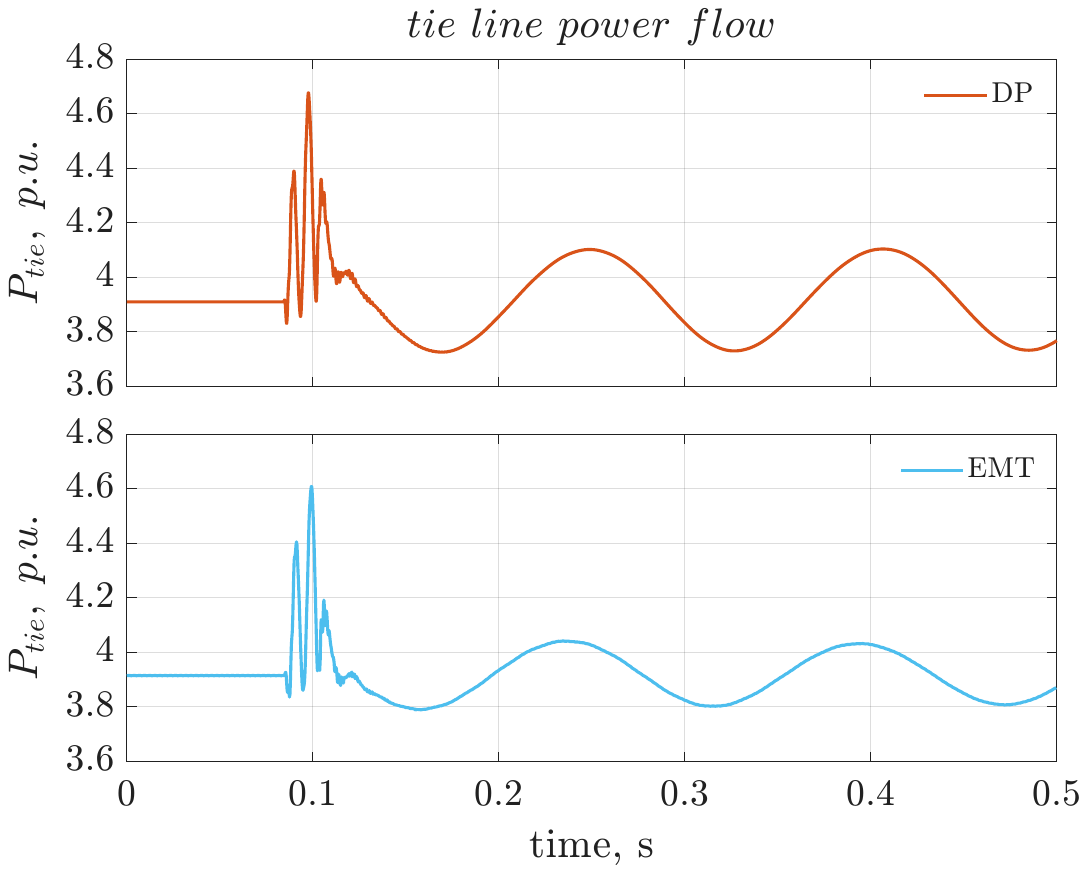}
	 \vspace{-17pt}
	\caption{Comparison of tie line power flow between DP and EMT models following a $1$ cycle self-clearing L-G fault near bus 10. ($20$-Hz PLL BW)}
	\label{fig:LG fault}
	\vspace{-10pt}
\end{figure}

\subsubsection{Validation of DP model of GFM IBR with current limits}
The DP-based GFM IBR model incorporating current-limiting strategies, shown in Fig. \ref{fig:GFC_SLD}, is validated by an EMTDC/PSCAD model, which is run with $1~\mu s$ timestep. Figures \ref{fig:GFM_DP_EMT}(a) and \ref{fig:GFM_DP_EMT}(b) show a close match between time domain responses in DP and EMT framework following a self-clearing L-L fault near bus $2$ for constant angle and $i_q$-priority current limiter, respectively. 

\subsubsection{Validation of DP-based system model with weak grid SSO induced by GFL IBRs}
For further validation of the approach in a multi-machine system incorporating GFL IBRs, we have performed both frequency- and time-domain analyses for the modified IEEE-$4$ machine system \cite{hossain2025dynamicphasorframeworkanalysis} as shown in Fig. \ref{fig:IEEE 4-machine system} where the SGs at bus $1$ and $2$ are replaced with two GFL IBRs of the corresponding ratings. The response of the DP model is compared to the output of a corresponding model built in EMTDC/PSCAD \cite{pscad_emtdc_v5}, which is run with $20~ \mu s$ timestep. The damping and frequency of the poorly-damped SSO mode are compared between models based on EMT (using Prony analysis \cite{prony1795}) and DP (using linearization) frameworks for two PLL bandwidths (BWs) $15$ and $20$ Hz as shown in Table \ref{tab:SSO_comparison}. \textcolor{black}{Participation factor analysis confirms that the SSO modes are induced by $\delta_{pll}$ of the GFL IBRs.} Figure \ref{fig:LG fault} shows the tie-line power flow between buses $7$ and $8$ following a self-clearing L-G fault near bus $10$ for PLL BW $20$ Hz. 

Minor discrepancies are observed between the DP and EMT simulation results, which can be attributed to variations in the exciter and governor representations of the synchronous generators. Additionally, differences arise from the manner in which $|v_{dq}|$ in Fig.~\ref{fig:GFL IBR_control}(b) is computed within the DP and EMT formulations. Moreover, the DP model employs a lumped $\pi$-equivalent transmission line representation, whereas the EMT model utilizes a Bergeron-based approach. It is also important to note that the identified SSO mode may exhibit sensitivity to parameter selection in the Prony analysis.

    
    

\begin{figure}
	\centering
	\includegraphics[width= 0.45\textwidth]{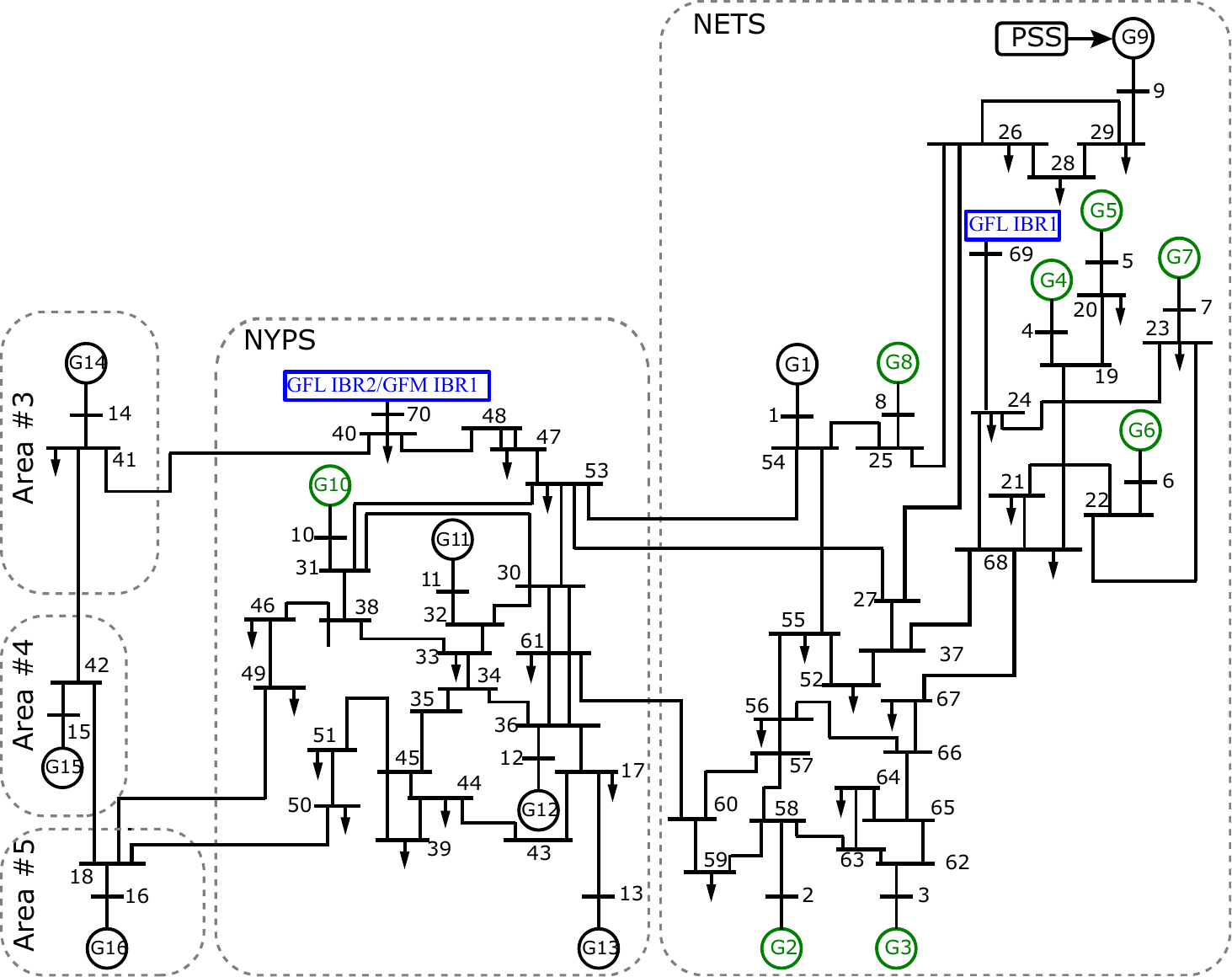}
	 \vspace{-8pt}
	\caption{Modified IEEE 68-bus system \cite{Ameli-25-TPWRS}.}
	\label{fig:IEEE-68 bus system}
	\vspace{-10pt}
\end{figure}

\begin{table}[!t]
\vspace{-10pt}
\caption{ Subsynchronous Oscillatory Modes}
\label{tab:sso_modes}
\centering
\begin{tabular}{|c|c|c|c|}
\hline
Mode & Eigenvalue, $\lambda$ & Damping Ratio, $\zeta$ & Frequency, Hz \\
\hline

1 &
$-0.529 \pm j12.0$ &
0.044 &
1.91 \\

2 &
$-7.52 \pm j18.9$ &
0.370 &
3.01 \\

3 &
$-0.070 \pm j29.9$ &
0.002 &
4.76 \\

4 &
$-21.9 \pm j47.7$ &
0.417 &
7.59 \\

5 &
$-293 \pm j203$ &
0.822 &
32.3 \\

6 &
$-304 \pm j203$ &
0.832 &
32.3 \\

\hline
\end{tabular}
\vspace{-10pt}
\end{table}
\begin{figure}[!t]
	\vspace{-0pt}
	\centering
	\includegraphics[width= 0.48\textwidth]{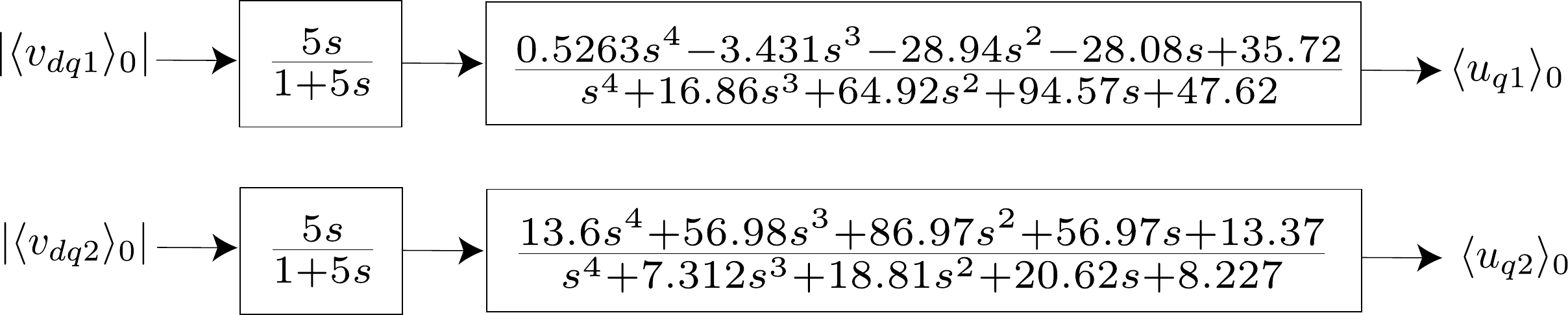}
	 \vspace{-5pt}
	\caption{Decentralized controller block diagram.}
	\label{fig:PSO_controller}
	\vspace{-10pt}
\end{figure}
\begin{figure}[!t]
	\centering
	\includegraphics[trim = {1cm 6.3cm 1.6cm 5.9cm}, clip,width= 0.4\textwidth]{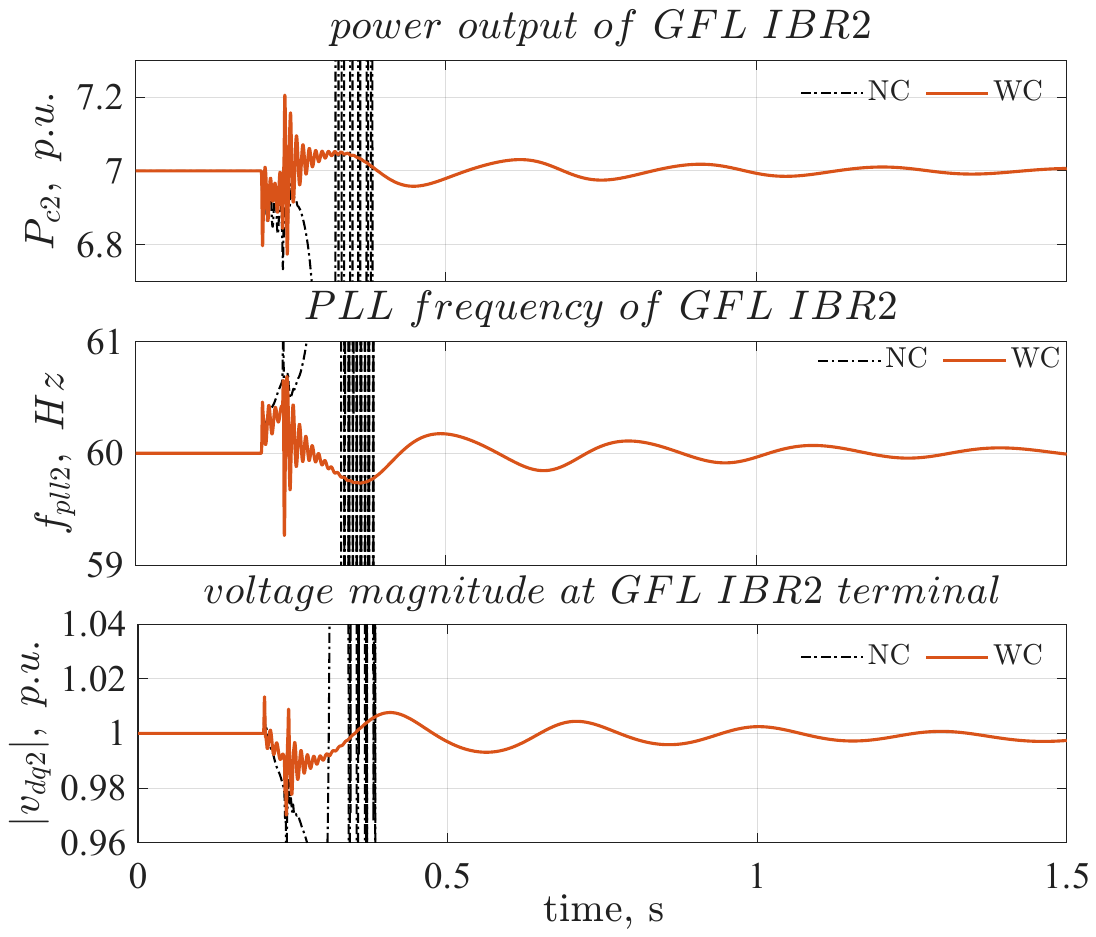}
	 \vspace{-10pt}
	\caption{Responses following a $2$ cycles self-clearing L-G fault near bus 50 while one of the parallel tie lines between bus 18 and bus 49 are out for maintenance. (NC: no damping control, WC: with damping control)}
	\label{fig:case_18_49}
	\vspace{-10pt}
\end{figure}

\begin{figure}[!t]
	\vspace{-5pt}
	\centering
	\includegraphics[trim = {1cm 5.5cm 1.6cm 5.5cm}, clip,width= 0.4\textwidth]{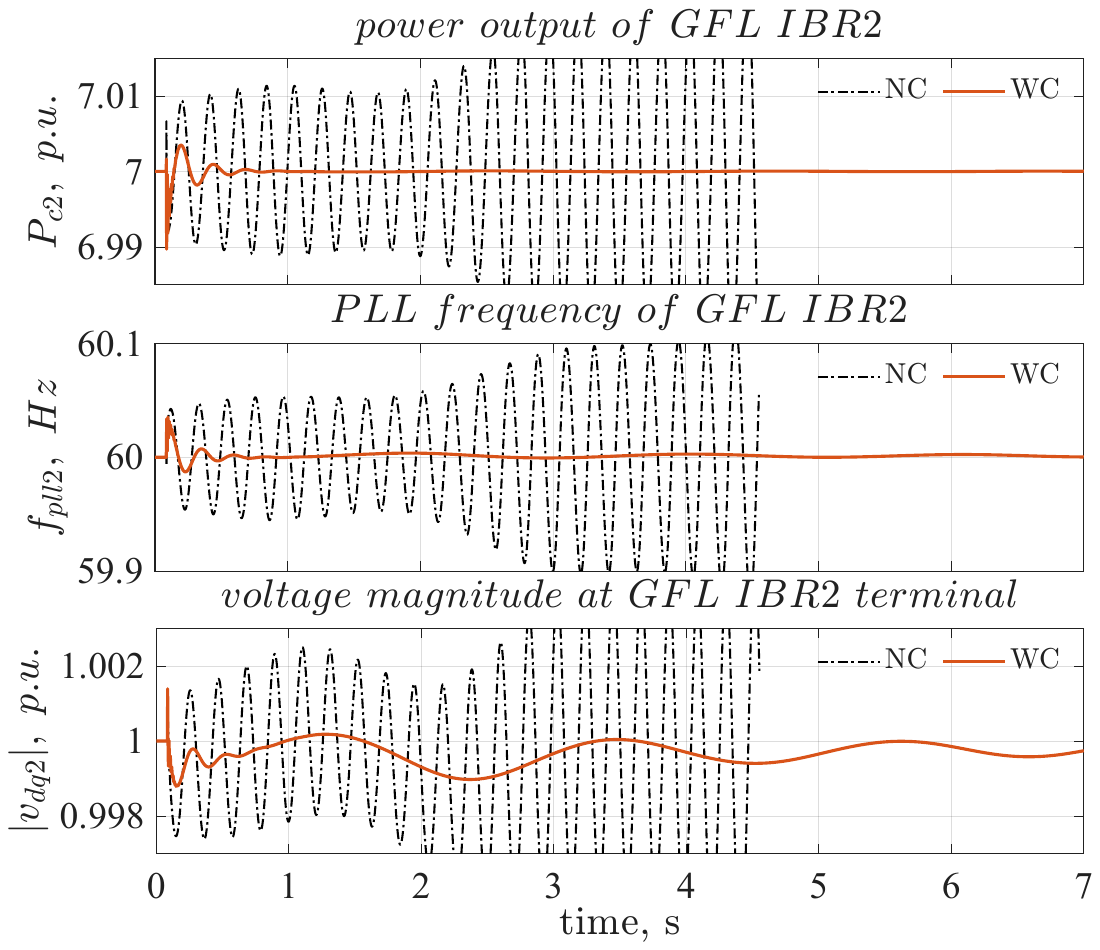}
	 \vspace{-17pt}
	\caption{Responses after one of the parallel tie line outage in $5$ cycles between bus $27$ and bus $53$ following a permanent L-G fault near bus $27$. (NC: no damping control, WC: with damping control)}
	\label{fig:case_27_53}
	\vspace{-20pt}
\end{figure}
\subsection{Damping of the poorly-damped SSO mode}
After gaining confidence in the DP models through validation, we built a DP model of the modified IEEE 68-bus system where two GFL IBRs are connected to buses $24$ and $40$ as shown in Fig. \ref{fig:IEEE-68 bus system}. Some oscillatory modes in the subsynchrnous frequency range are reported in Table \ref{tab:sso_modes}. A poorly-damped SSO mode with frequency $4.76$ Hz and damping ratio $0.2\%$ is present in the system, \textcolor{black}{which can be attributed to $\delta_{pll}$ of GFL IBR2 as confirmed by participation factor analysis.} We present two solutions to damp the SSO mode.
\begin{figure}
	\vspace{-9pt}
	\centering
	\includegraphics[trim = {1cm 5cm 1.6cm 5.5cm}, clip,width= 0.4\textwidth]{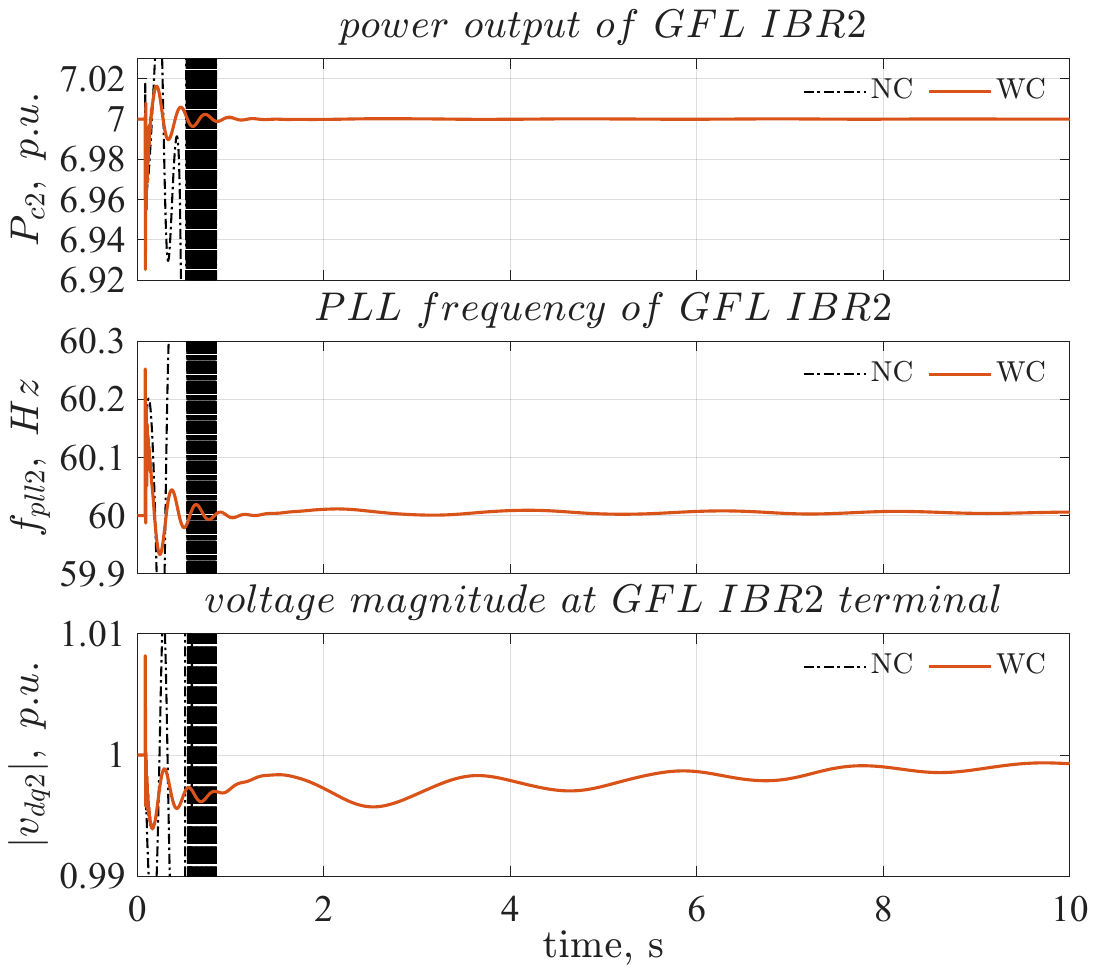}
	 \vspace{-18pt}
	\caption{Responses after one of the parallel tie line outage in $3$ cycles between bus $53$ and bus $54$ following a permanent L-G fault near bus $53$. (NC: no damping control, WC: with damping control)}
	\label{fig:case_53_54}
	\vspace{-10pt}
\end{figure}
\begin{table}[!t]
\vspace{-1pt}
\centering
\caption{Subsynchronous Oscillatory Modes}
\label{tab:sso_modes_modified}
\begin{tabular}{|c|c|c|c|}
\hline
Mode & Eigenvalue, $\lambda$ & Damping Ratio, $\zeta$ & Frequency, Hz \\
\hline

1 &
$-15.9 \pm j11.7$ &
0.806 &
1.86 \\

2 &
$-21.9 \pm j47.6$ &
0.418 &
7.57 \\

3 &
$-304 \pm j203$ &
0.832 &
32.3 \\

4 &
$-38.5 \pm j351$ &
0.109 &
55.9 \\

\hline
\end{tabular}
\vspace{-5pt}
\end{table}
\begin{figure}[!t]
	\vspace{-15pt}
	\centering
	\includegraphics[trim = {1cm 6cm 1.6cm 5cm}, clip,width= 0.4\textwidth]{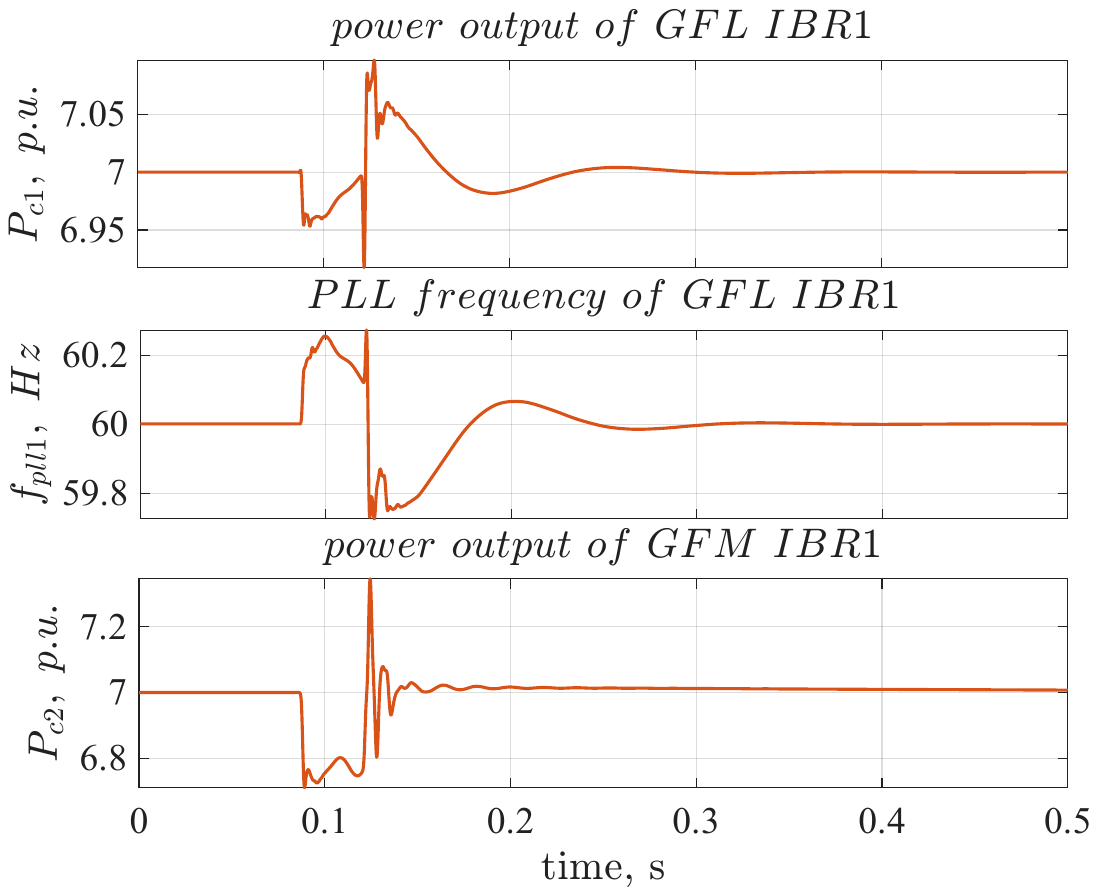}
	 \vspace{-20pt}
	\caption{Responses following a $2$ cycles self-clearing LLL-G fault near bus 50 (GFL IBR2 is replaced with GFM IBR1).}
	\label{fig:GFM_IBR_response}
	\vspace{-20pt}
\end{figure}
\subsubsection{Solution I: Decentralized damping controller design} 
As described in our previous work \cite{Chaudhuri-13-DFIG-PSO} the design problem of a low order, decentralized, fixed structure, and stable supplementary control for oscillation damping is non-convex. We have used a heuristic method called particle swarm optimization to design the controllers \cite{Chaudhuri2009RobustPOD} considering the operating condition, where one of the parallel tie lines between buses $18$ and $49$ is out for maintenance. The designed $5$th order controller, shown in Fig. \ref{fig:PSO_controller}, uses $|\langle v_{dq}\rangle_0|$ of each GFL as feedback signals and modulates their respective reactive current references $\langle i_{tq}^*\rangle_0$ (Fig., \ref{fig:GFC_control}(b)) to achieve a settling time of at most $15$ s for all modes. 

Figures \ref{fig:case_18_49} -- \ref{fig:case_53_54} show the effectiveness of controller performance following one self-clearing L-G fault and multiple L-G faults followed by various tie-line outages. Note that the IBR SSO mode is effectively damped in each case. The three clear distinctions between our previous paper \cite{Ameli-25-TPWRS} and this work are -- (1) \cite{Ameli-25-TPWRS} did not consider decentralized design involving multiple IBRs, (2) the modeling framework in \cite{Ameli-25-TPWRS} did not allow unbalanced fault simulation, (3) \cite{Ameli-25-TPWRS} modulates the real power reference while this paper considers the reactive power channel for control input that allows maximum power point tracking.


\subsubsection{Solution II: Replacing GFL IBR with GFM IBR}\label{sub:GFM IBR}
The poorly damped SSO mode disappears if GFL IBR2 in Fig. \ref{fig:IEEE-68 bus system} is replaced by GFM IBR1 of the same rating; see Table \ref{tab:sso_modes_modified}. Figure \ref{fig:GFM_IBR_response} shows the time domain responses following a self-clearing LLL-G fault near bus 50. It further validates the disappearance of the poorly damped SSO mode.

\begin{figure}[!t]
	\vspace{-5pt}
	\centering
	\includegraphics[trim = {1.1cm 9.5cm 1.6cm 6cm}, clip,width= 0.38\textwidth]{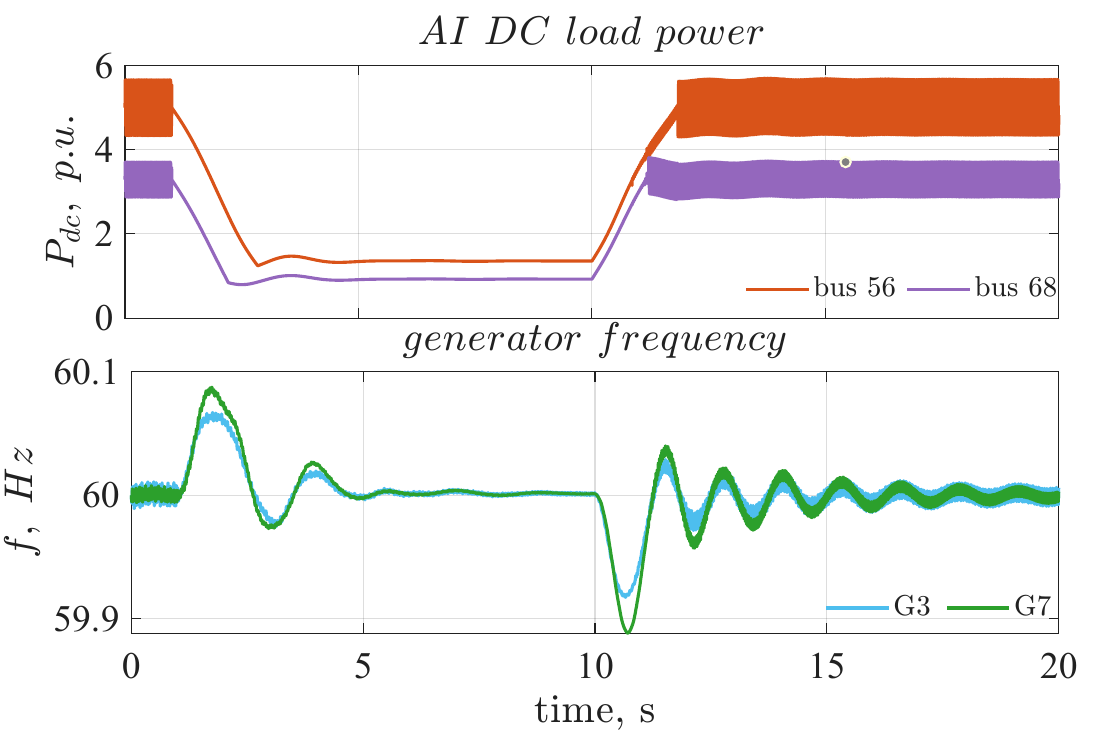}
	 \vspace{-10pt}
	\caption{Primary frequency response following AI DC load ramp.}
	\label{fig:primary_frequency_response}
	\vspace{-6pt}
\end{figure}
\begin{figure}
\vspace{-5pt}
    \centering
    
    \begin{minipage}{0.38\textwidth}
        \centering
        \includegraphics[trim={1.3cm 5.5cm 1.6cm 5.5cm},clip,width=\textwidth]{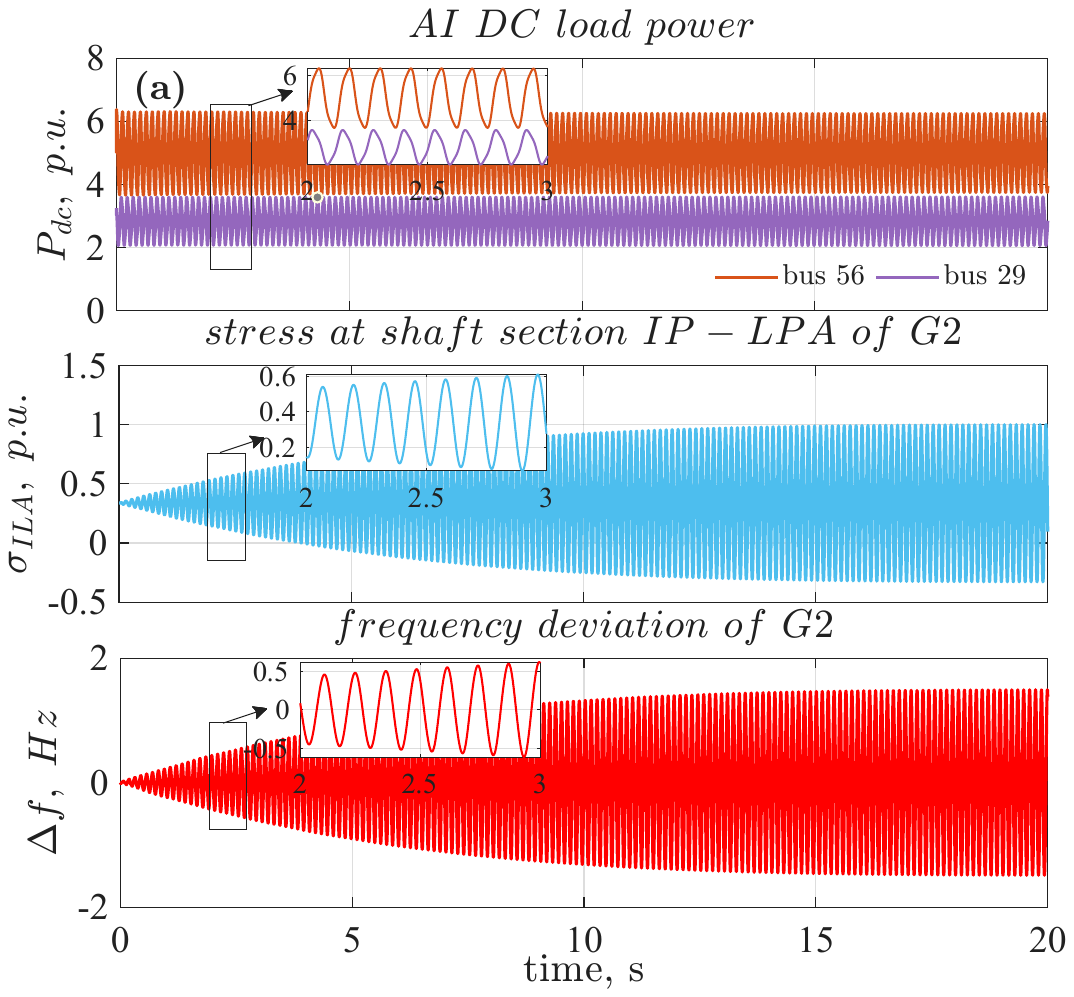}
        \vspace{-0.5cm}
        \label{fig:time_domain}
    \end{minipage}
    
    
    \begin{minipage}{0.38\textwidth}
        \centering
        \includegraphics[trim={0.8cm 5.5cm 1.6cm 6cm},clip,width=\textwidth]{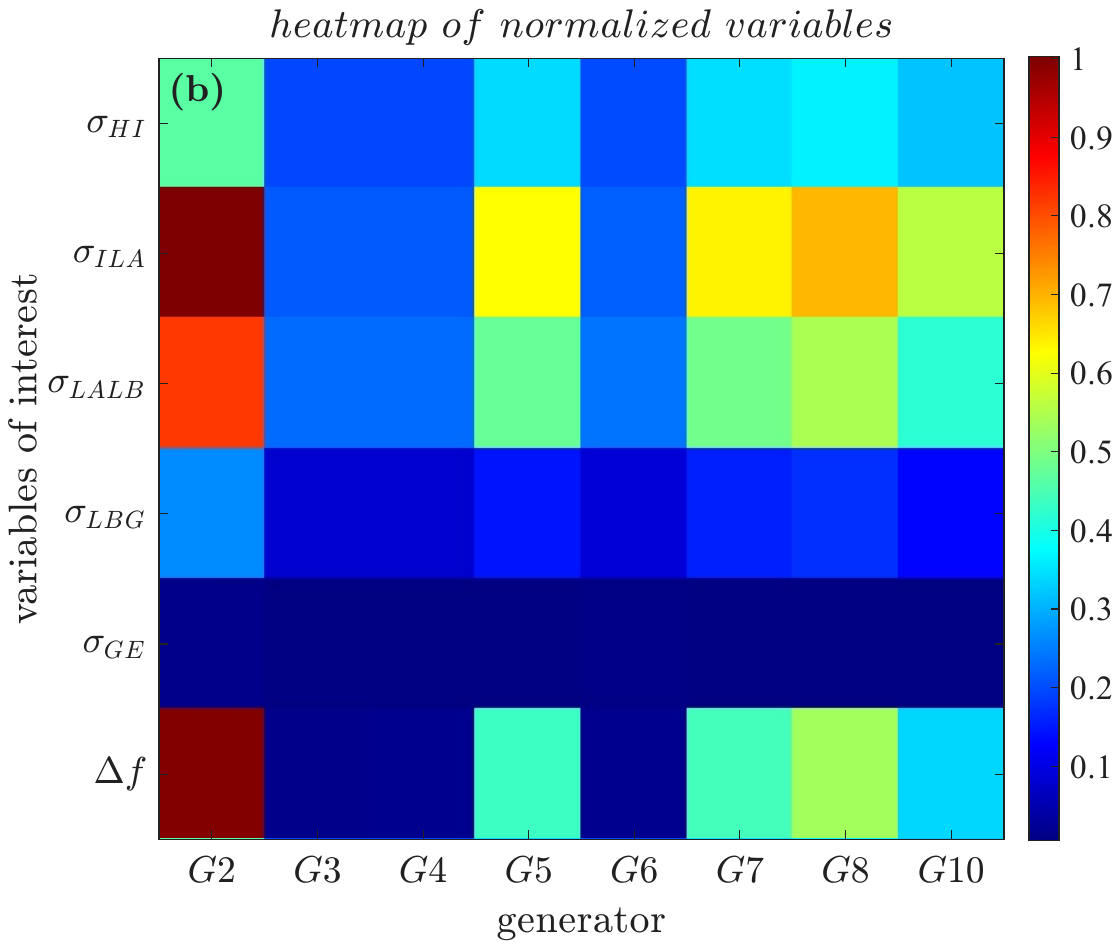}
        \vspace{-0.7cm}
        \label{fig:heatmap}
    \end{minipage}
    
    \caption{Time-domain response and corresponding heatmap for the considered case.}
    \label{fig:time_heatmap}
    \vspace{-18pt}
\end{figure}

\begin{figure*}[t]
    \centering

    \begin{minipage}{0.32\textwidth}
        \centering
        \includegraphics[trim={1.1cm 5.5cm 1.6cm 6cm},clip,width=\textwidth]
        {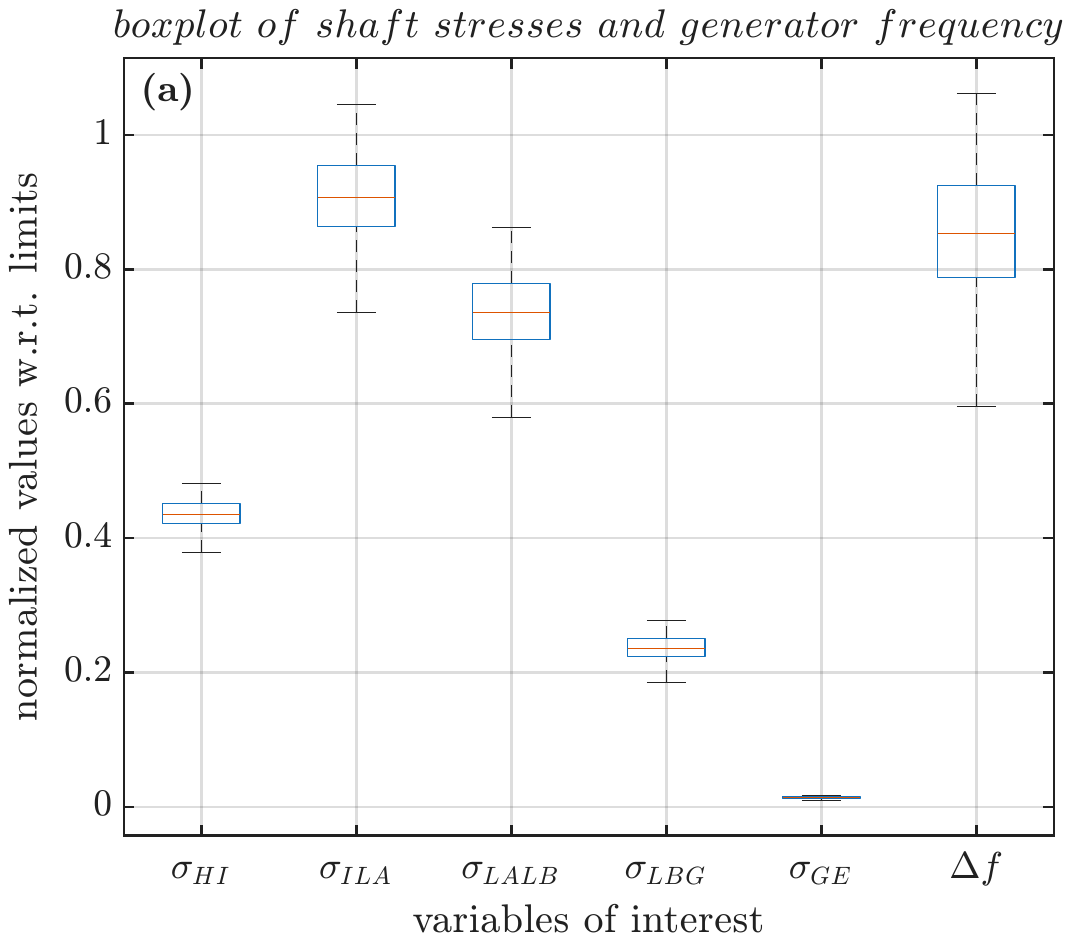}
    \end{minipage}
    \hfill
    \begin{minipage}{0.32\textwidth}
        \centering
        \includegraphics[trim={1.1cm 5.5cm 1.6cm 6cm},clip,width=\textwidth]
        {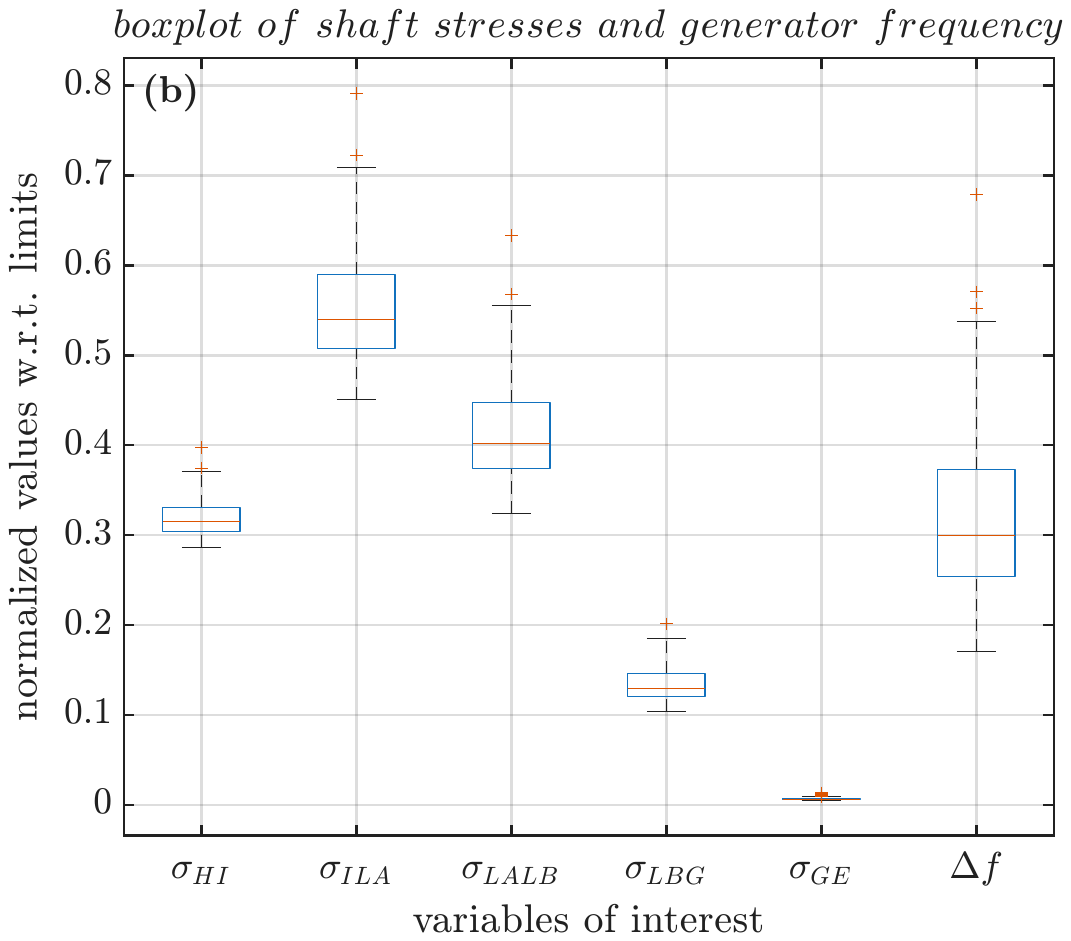}
    \end{minipage}
    \hfill
    \begin{minipage}{0.32\textwidth}
        \centering
        \includegraphics[trim={1.1cm 5.5cm 1.6cm 6cm},clip,width=\textwidth]
        {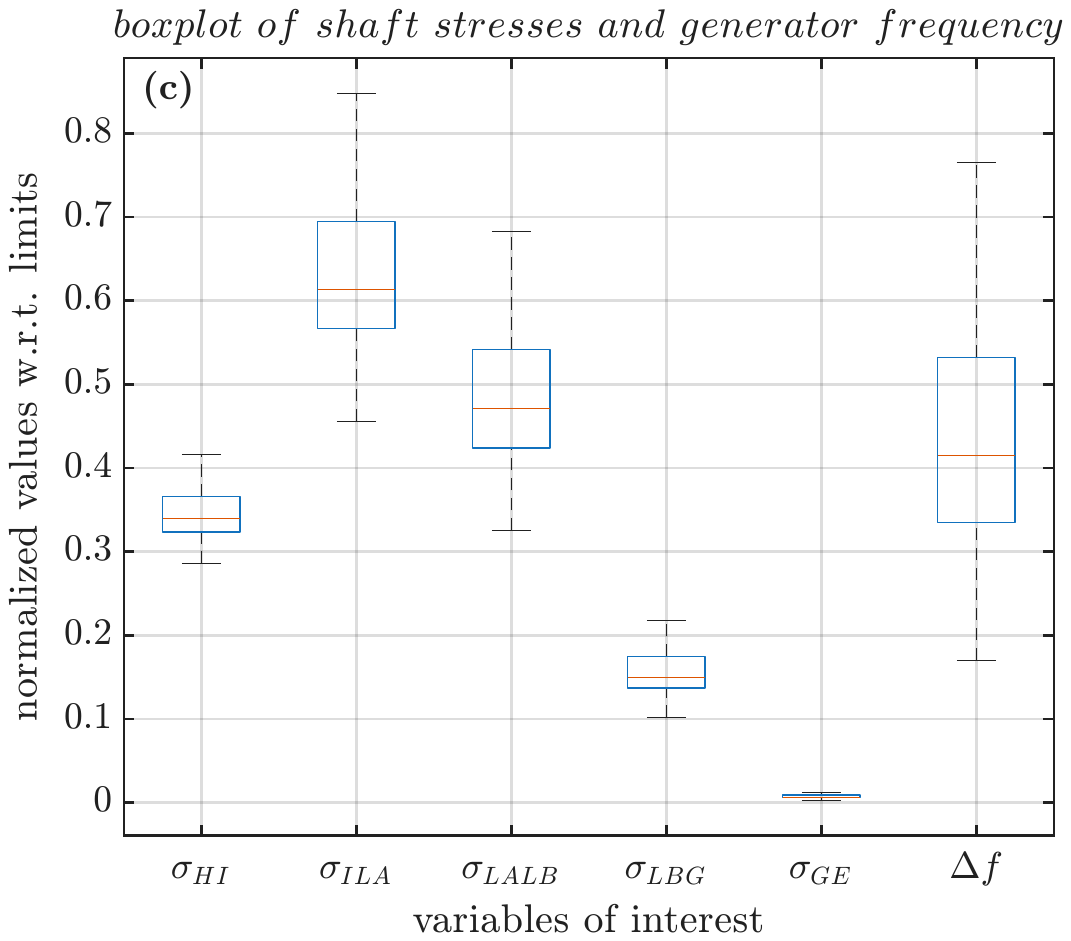}
    \end{minipage}
    \vspace{-7pt}
    \caption{
    Boxplots of normalized shaft stresses and generator frequency with respect to their corresponding limits (maximum values considered) for 100 Monte Carlo simulations during the compute phase. The limits for shaft stresses are obtained from the augmented modified Goodman diagram. The simulations are performed by randomly varying (a) oscillation amplitudes, (b) oscillation frequencies, and (c) oscillation phases.
    }
    \vspace{-15pt}\label{fig:boxplot_montecarlo}
\end{figure*}

\vspace{-15pt}
\subsection{Impact of AI DC loads on primary frequency response and SG turbine shaft stress}
The generators G2-G8 and G10 are modeled with multi-mass turbines in the modified system of subsection \ref{sub:GFM IBR}. Generator G1 is modeled as a hydroelectric unit. The local oscillatory mode of G9 is damped using a power system stabilizer (PSS); redesigning this PSS for the detailed multi-mass shaft representation was intentionally avoided. The other generators correspond to equivalent or aggregated representations of areas and are therefore not equipped with multi-mass turbine models. The shaft system parameters employed for generators G3, G4, and G6, as well as for G2, G5, G7, G{8}, and G{10}, are adopted from \cite{IEEESSRBenchmark1977} and \cite{kundur1994power}, respectively. Since most of the multi-mass turbine equipped SGs are located in NETS, AI DC loads are deliberately chosen at buses 56, 68, 55, 67, 24, 29, and 59. The primary frequency response due to changes in AI DC loads, transitioning from the compute period to the rest period at $t=1~s$ and from the rest period to the compute period at $t=10~s$ are demonstrated in Fig. \ref{fig:primary_frequency_response}. The frequency variation in this case is large enough to trigger frequency support from ancillary resources, but will not lead to load shedding or generator tripping. 

To show the impact of AI DC load variation during the compute phase on turbine shafts and blades, the power oscillation frequency is deliberately chosen around $7.83$ Hz (coinciding with one of the torsional modes) along with its third harmonic. If the oscillation amplitudes are $28.71\%$ of their respective nominal loads and their phases are such that the load variations contribute in phase to the shaft stress at the IP--LPA section of G2, the resulting stress reaches the no-fatigue-life-loss limit determined from the augmented modified Goodman diagram. Figure \ref{fig:time_heatmap}(a) shows the time domain response of stress and frequency of the generators due to such AI DC load variations. Normalized shaft stresses and frequencies for all the multi-mass turbine SGs with respect to the limit obtained from augmented modified Goodman diagram are presented in Fig. \ref{fig:time_heatmap}(b). Please see reference \cite{hossain2026limitingimpactaidata} for more details.

To determine the sensitivity of shaft stresses to variations in oscillation amplitude, frequency, and phase of the AI DC load, $100$ Monte Carlo simulations are performed by randomly changing each variable individually while keeping the others fixed. The amplitudes are varied from $15\%$ to $35\%$ of the respective nominal loads, the frequencies from $7.63$ to $8.03$ Hz, and the phases from $0$ to $2\pi$. The resulting statistical distributions of the shaft stress indices are presented as box plots in Fig. \ref{fig:boxplot_montecarlo}. The median values indicate that shaft stresses and $\Delta f$ can decrease significantly even with a small change in the oscillation frequency from the torsional mode. The sensitivity of the median to the phase angle is less dominant, but the inter-quartile range is larger compared to the amplitude variation.
\vspace{-15pt}

\begin{table}[!t]
\caption{IEEE 4-machine system: Runtime stat of $100$ simulations for $20$ s}
\label{tab:runtime}
\centering
\resizebox{0.40\textwidth}{!}{
\begin{tabular}{|c|c|c|c|c|}
\hline
\textbf{Framework} & \textbf{Max (s)} & \textbf{Min (s)} & \textbf{Mean (s)} & \textbf{Std.  dev. (s)}\\ \hline
EMT$^*$ (Fig.\ref{fig:LG fault}) & 113.41 & 96.658 & 103.871 & 5.545 \\ \hline
DP (Fig.\ref{fig:LG fault}) & 9.445 & 7.735 & 8.168 & 0.333\\ \hline

\end{tabular}
}
\vspace{-5pt}
\small
\begin{flushleft}
\hspace{30pt}\footnotesize \textsuperscript{*} averaged model of IBRs used for fair comparison 
\end{flushleft}
\normalsize
\vspace{-15pt}
\end{table}
\begin{table}[!t]
\vspace{-4pt}
\caption{IEEE 68-bus system: Runtime stat of $100$ simulations for $20$ s}
\label{tab:runtime_AI_DC}
\centering
\resizebox{0.40\textwidth}{!}{
\begin{tabular}{|c|c|c|c|c|}
\hline
\textbf{Case study} & \textbf{Max (s)} & \textbf{Min (s)} & \textbf{Mean (s)} & \textbf{Std.  dev. (s)}\\
\hline
 Fig. \ref{fig:case_18_49} (WC)& 112.74 & 91.42 & 98.52 & 4.06\\ \hline
  Fig. \ref{fig:case_27_53} (WC)& 535.24 & 423.37 & 451.16 & 39.73\\ \hline
 Fig. \ref{fig:boxplot_montecarlo}(c)& 379.25 & 124.65 & 154.43 & 38.60\\ \hline

\end{tabular}
}
\vspace{-15pt}
\end{table}
\subsection{Runtime statistics}
Statistical analysis of $100$ simulations for the case of Fig. \ref{fig:LG fault} is presented in Table \ref{tab:runtime} for EMT and DP models. The DP based model runs $12$x times faster than the EMT model on average. Finally, we present the simulation runtime statistics of the modified IEEE 68-bus system model in DP framework in Table~\ref{tab:runtime_AI_DC}. Three types of cases were simulated for $20$ s each and $100$ Monte Carlo runs were considered for each case. The first two cases involve simulation with the SSO damping controller following unbalanced faults (Figs~\ref{fig:case_18_49}, \ref{fig:case_27_53}). The average runtime for the 5-cycle fault followed by line outage is $\approx 7.5$ min, which is much larger compared to the 2-cycle self-clearing fault simulation. The third case simulating periodic large loads consume $\approx 2.5$ min on average. These run times are quite reasonable for the modified IEEE $68$-bus system and establish scalability of the DP framework.

\vspace{-10pt}
\section{Conclusions}
To the best of our knowledge, this is the first work that presented a DP framework for simulating multi-machine power systems incorporating GFL and GFM IBRs, multi-mass turbine-generators, and transmission line dynamics. Extensive validation of different aspects, such as torsional interaction, current limiting in IBR under unbalanced faults, and IBR-induced SSOs with EMT models, has been performed. The utility of the time-invariance of the proposed framework for control design using linearization has been demonstrated. It has been shown that stresses in the turbine-generator shafts can be calculated in the presence of fluctuations in AI DC loads in a modified IEEE 68-bus system using the proposed model in a reasonable time. Although the system considered is highly stiff consisting of 8 SGs with multi-mass models, 8 SGs with single-mass model, 1 GFL IBR, 1 GFM IBR, and dynamic model of transmission lines; it took on average only 2.5 minutes to simulate a forced oscillation phenomenon from 7 AI DCs for 20 s.              
\vspace{-10pt}

\bibliographystyle{IEEEtran}
\bibliography{Mybib}

@article{Chaudhuri2009RobustPOD,
  author    = {Balarko Chaudhuri and Swakshar Ray and Rajat Majumder},
  title     = {Robust low-order controller design for multi-modal power oscillation damping using flexible AC transmission systems devices},
  journal   = {IET Generation, Transmission \& Distribution},
  year      = {2009},
  volume    = {3},
  number    = {5},
  pages     = {448--459},
  doi       = {10.1049/iet-gtd.2008.0471},
  publisher = {Institution of Engineering and Technology (IET)}
}

@ARTICLE{Chaudhuri-13-DFIG-PSO,
  author={Chaudhuri, Nilanjan Ray and Chaudhuri, Balarko},
  journal={IEEE Transactions on Power Delivery}, 
  title={Considerations Toward Coordinated Control of DFIG-Based Wind Farms}, 
  year={2013},
  volume={28},
  number={3},
  pages={1263-1270},
  doi={10.1109/TPWRD.2013.2263429}}

@article{prony1795,
  author    = {Gaspard Riche de Prony},
  title     = {Essai expérimental et analytique: Sur les lois de la dilatabilité des fluides élastiques et sur celles de la force expansive de la vapeur de l'eau et de la vapeur de l'alkool, à différentes températures},
  journal   = {Journal de l'École Polytechnique},
  year      = {1795},
  volume    = {1},
  number    = {22},
  pages     = {24--76}
}

@misc{pscad_emtdc_v5,
  author       = {{Manitoba Hydro International Ltd.}},
  title        = {{PSCAD/EMTDC\texttrademark\ Version 5.0}},
  year         = {2024},
  note         = {Computer software},
  url          = {https://www.pscad.com}
}

@INPROCEEDINGS{Demiray-08-DP-39bus,
  author={Demiray, T. and Andersson, G. and Busarello, L.},
  booktitle={2008 IEEE/PES Transmission and Distribution Conference and Exposition: Latin America}, 
  title={Evaluation study for the simulation of power system transients using dynamic phasor models}, 
  year={2008},
  volume={},
  number={},
  pages={1-6},
  doi={10.1109/TDC-LA.2008.4641716}}

@misc{hossain2026limitingimpactaidata,
      title={Limiting the Impact of AI Data Centers on Fatigue Life of Thermal Turbine Generators in the Grid: A Frequency-Domain Approach}, 
      author={Fiaz Hossain and Nilanjan Ray Chaudhuri and Alok Sinha and Sai Gopal Vennelaganti and Mohammed E. Nassar},
      year={2026},
      eprint={2605.01173},
      archivePrefix={arXiv},
      primaryClass={eess.SY},
      url={https://arxiv.org/abs/2605.01173}, 
}

@misc{HossainChaudhuriLagoa2026DPSSO,
  title         = {A Dynamic Phasor Framework for Analysis of {IBR}-Induced {SSOs} in Multi-Machine Systems},
  author        = {Hossain, Fiaz and others},
  year          = {2026},
  eprint        = {2604.21234},
  archivePrefix = {arXiv},
  primaryClass  = {eess.SY},
  doi           = {10.48550/arXiv.2604.21234},
  url           = {https://arxiv.org/abs/2604.21234}
}

@ARTICLE{Verghese-91-DP,
  author={Sanders, S.R. and Noworolski, J.M. and Liu, X.Z. and Verghese, G.C.},
  journal={IEEE Trans. on Power Electronics}, 
  title={Generalized averaging method for power conversion circuits}, 
  year={1991},
  volume={6},
  number={2},
  pages={251-259},
  doi={10.1109/63.76811}}

@ARTICLE{Stankovic_assymetric_SMIB,
  author={Stankovic, A.M. and Aydin, T.},
  journal={IEEE Trans. on Power Syst.}, 
  title={Analysis of asymmetrical faults in power systems using dynamic phasors}, 
  year={2000},
  volume={15},
  number={3},
  pages={1062-1068},
  doi={10.1109/59.871734}}

@ARTICLE{DP_SSR,
  author={Chudasama, Mahipalsinh C. and Kulkarni, Anil M.},
  journal={IEEE Trans. on Power Syst.}, 
  title={Dynamic Phasor Analysis of {SSR} Mitigation Schemes Based on Passive Phase Imbalance}, 
  year={2011},
  volume={26},
  number={3},
  pages={1668-1676},
  doi={10.1109/TPWRS.2010.2072793}}

@INPROCEEDINGS{hossain2025dynamicphasorframeworkanalysis,
  author={Hossain, Fiaz and Chaudhuri, Nilanjan Ray},
  booktitle={2025 IEEE Power \& Energy Society General Meeting (PESGM)}, 
  title={A Dynamic Phasor Framework for Analysis of Grid-Forming Converter Connected to Series-Compensated Line}, 
  year={2025},
  volume={},
  number={},
  pages={1-5},
  doi={10.1109/PESGM52009.2025.11225603}}

@ARTICLE{Vega_Herrera,
  author={Vega-Herrera, Jorge and Rahmann, Claudia and Valencia, Felipe and Strunz, Kai},
  journal={IEEE Trans. on Power Syst.}, 
  title={Analysis and Application of Quasi-Static and Dynamic Phasor Calculus for Stability Assessment of Integrated Power Electric and Electronic Systems}, 
  year={2021},
  volume={36},
  number={3},
  pages={1750-1760},
  doi={10.1109/TPWRS.2020.3030225}}

@ARTICLE{YangTao,
  author={Yang, Tao and others},
  journal={IEEE Trans. on Power Syst.}, 
  title={Dynamic Phasor Modeling of Multi-Generator Variable Frequency Electrical Power Systems}, 
  year={2016},
  volume={31},
  number={1},
  pages={563-571},
  doi={10.1109/TPWRS.2015.2399371}}

@ARTICLE{Real_SSO_event,
  author={Cheng, Yunzhi and others},
  journal={IEEE Trans. on Power Syst.}, 
  title={Real-World Subsynchronous Oscillation Events in Power Grids With High Penetrations of Inverter-Based Resources}, 
  year={2023},
  volume={38},
  number={1},
  pages={316-330},
  doi={10.1109/TPWRS.2022.3161418}}

@techreport{NERC_2025,
  author      = {NERC},
  title       = {Characteristics and Risks of Emerging Large Loads},
  institution = {North American Electric Reliability Corporation},
  year        = {2025},
  type        = {White Paper},
}

@techreport{DC1A,
  author      = {{IEEE Power \& Energy Society}},
  title       = {{IEEE} Recommended Practice for Excitation System Models for Power System Stability Studies ({IEEE} {Std.} 421.5-2005)},
  institution = {IEEE},
  year        = {2005},
  month       = apr
}

@manual{matlab,
  address      = {Natick, Massachusetts},
  organization = {The MathWorks, Inc.},
  title        = {{MATLAB version 9.14.0.2286388 (R2023a)}},
  year         = {2023},
  }

@article{IEEESSRBenchmark1977,
  author  = {{IEEE Working Group}},
  title   = {First Benchmark Model for Computer Simulation of Subsynchronous Resonance},
  journal = {IEEE Trans. Power Appar. Syst.},
  volume  = {PAS-96},
  number  = {5},
  pages   = {1565--1572},
  year    = {1977},
  month   = sep,
  doi     = {10.1109/T-PAS.1977.32485}
}

@misc{Young2026SSR,
  author       = {Young, Meme},
  title        = {{IEEE First Benchmark on Subsynchronous Resonance (SSR)}},
  year         = {2026},
  howpublished = {MATLAB Central File Exchange},
  url          = {https://www.mathworks.com/matlabcentral/fileexchange/99539-ieee-first-benchmark-on-subsynchronous-resonance-ssr},
  note         = {Retrieved April 23, 2026}
}

@ARTICLE{Ameli-25-TPWRS,
  author={Ameli, Sina and others},
  journal={IEEE Trans. Power Syst.}, 
  title={Robust Adaptive Supplementary Control for Damping Weak-Grid {SSOs} Involving {IBRs}}, 
  year={2025},
  volume={40},
  number={5},
  pages={4322-4335},
  doi={10.1109/TPWRS.2025.3544541}}

@book{kundur1994power,
  author    = {Kundur, Prabha},
  title     = {Power System Stability and Control},
  publisher = {McGraw-Hill},
  address   = {New York},
  year      = {1994},
  isbn      = {978-0070359581}
}

\end{document}


\title{Supplementary Document
}

\author{Fiaz Hossain,~\IEEEmembership{Student Member,~IEEE},  Nilanjan Ray Chaudhuri,~\IEEEmembership{Senior Member,~IEEE}, Constantino M. Lagoa,~\IEEEmembership{Member,~IEEE}, Alok Sinha, Sai Gopal Vennelaganti,~\IEEEmembership{Member,~IEEE}, and Mohammed E. Nassar,~\IEEEmembership{Senior Member,~IEEE}}

\maketitle 
\section*{Nomenclature}
\begin{longtable}{ll}
superscript($^*$) & Reference quantities\\
$|.|$ & Magnitude of the quantities \\
$T_m$ & Generator mechanical input torque \\
$T_{gm}$ & Governor output torque \\
$T_{tm}$ & Turbine mechanical torque \\
$F_r$ & Fraction of mechanical input torque going to $r$th mass\\
$\delta_H,\delta_I,\delta_{LA}, \delta_{LB}, \delta_E$ & Rotor angle of HP, IP, LPA, LPB mass section of turbine, and exciter, respectively.\\
$\omega_H,\omega_I,\omega_{LA}, \omega_{LB}, \omega_E$ & Angular speed of HP, IP, LPA, LPB mass section of turbine, and exciter, respectively.\\
$\delta_g, \omega_g$ & Rotor angle and angular speed of generator\\
$\delta_r,\omega_r$ & Rotor angle and angular speed of $r$th mass\\
$\psi_{pnz}=[\psi_p~\psi_n~\psi_z]^T$ & Positive, negative, and zero sequence stator flux linkage\\
$\psi_r =[\psi_f~\psi_h~\psi_g~\psi_k]^T$ & Field and damper winding flux linkage \\
$V_r$ & AVR regulator output voltage \\
$R_f$ & Exciter stabilizing feedback state\\
$V_{tr}$ & Transducer (voltage sensor) output \\
$e_{fd}$ & field voltage applied to the generator field winding\\
$K_{p,ac},K_{i,ac}$ & Proportional and integral gain of outer voltage controller of GFM IBR \\
$\tau_p$ & Power measurement delay of GFM IBR \\
$d_{pc}$ & Droop constant of GFM IBR\\
$\omega_s$ & Nominal angular frequency \\
$\omega_c$ & Converter angular frequency \\
$\omega_{pll}$ & PLL angular frequency \\
$k_{vp},k_{vi}$ & Proportional and integral gain of inner voltage controller of GFM IBR \\
$k_{cp},k_{ci}$ & Proportional and integral gain of current controller of GFM IBR \\
$k_p,k_i$ & Proportional and integral gain of PLL controller \\
$\tau_m,\tau_f$ & Voltage measurement and feedback delay of GFL IBR \\
$k_{p,v},k_{i,v}$ & Proportional and integral gain of outer control loop of GFL IBR \\
$k_{p,i},k_{i,i}$ & Proportional and integral gain of inner current control loop of GFL IBR \\
$R,L,C$ & Filter resistance, inductance, and capacitance of GFM and GFL IBR \\
$R_t, L_t$ & Transformer resistance and inductance \\
$H_r$ & Inertia constant of $r$th mass \\
$K_r$ & Stiffness coefficient of shaft between $r$th and $(r+1)$th mass \\
$D_r$ & Mutual damping coefficient of shaft between $r$th and $(r+1)$th mass \\
$\bar{D}_r$ & Self damping coefficient of $r$th mass \\
$G$ & Shear modulus of elasticity \\
$R_r$ & Radius of $r$th shaft \\
$l_r$ & Length of $r$th shaft \\
$\sigma_{HI}, \sigma_{ILA},\sigma_{LALB}, \sigma_{LBG}, \sigma_{GE}$ & Stress at shaft section HP-IP, IP-LPA, LPA-LPB, LPB-Gen, and Gen-Exc, respectively\\
$R_{sp}$ & Speed regulation constant \\
$\tau_g,\tau_t$ & Governor and turbine time constant \\
$[R_s] = diag(R_a,R_a,R_a)$ & Stator winding resistances \\
$[R_r] = diag(R_f,R_h,R_g,R_k)$ & Rotor winding resistances \\
$L_{aa0}, L_{aa2}$ & Constant and rotor position dependent portion of stator phase a winding self-inductance \\
$L_{ab0}$ & Constant mutual inductance between stator winding phases\\
$M_{af}, M_{ah}, M_{ag}, M_{ak}$ & Mutual inductance between stator phase a winding and rotor windings \\
$L_f, L_h,L_g, L_k$ & Self-inductance of rotor windings \\
$L_{fh},L_{gk}$ & Mutual inductance between same axis rotor windings \\
$L_{adu}$ & Unsaturated $d$-axis mutual inductance between stator and rotor circuits\\
$T_r$ & Transducer time constant \\
$K_A, T_A$ & Regulator gain and time constant \\
$K_F, T_F$ & Rate feedback gain and time constant \\
$K_E, T_E$ & Exciter gain and time constant \\
$A_{ex}, B_{ex}$ & Exponential saturation function coefficient \\
$T_s$ & Speed measurement sensor delay \\
$T_w$ & Washout time constant \\
$T_1,T_2, T_3, T_4$ & Lead-lag time constant \\
$K_{PSS}$ & Power system stabilizer gain \\
$[R_l],[L_l],[C_l]$ & Transmission line resistance, inductance, and capacitance matrix \\
$[R_L], [L_L]$ & Load resistance and inductance matrix \\
$v_{cd},v_{cq}$ & $d$-axis and $q$-axis component of voltage across filter capacitor of GFM IBR \\
$P_c,Q_c$ & Active and reactive power output of GFL/GFM IBR \\
$\tilde{P_c}$ & Filtered real power feedback signal of GFM IBR \\
$\tilde{P_c}$ & Filtered real power feedback signal of GFM IBR \\
$\delta_c$ & Angle between asynchronous $dq$ frame of GFM IBR and synchronous $DQ$ frame \\
$\delta_{pll}$ & Angle between asynchronous $dq$ frame of GFL IBR and synchronous $DQ$ frame \\
$v_{td},v_{tq}$ & $d$-axis and $q$-axis component of inverter terminal voltage of GFL/GFM IBR \\
$i_{td},i_{tq}$ & $d$-axis and $q$-axis component of filter $RL$ branch current of GFL/GFM IBR \\
$i_{d},i_{q}$ & $d$-axis and $q$-axis component of output current of GFL/GFM IBR \\
$x_{1d},x_{1q}$ & $d$-axis and $q$-axis component of inner voltage controller state of GFM IBR \\
$x_{2d},x_{2q}$ & $d$-axis and $q$-axis component of current controller state of GFM IBR \\
$x_{3d},x_{3q}$ & $d$-axis and $q$-axis component of filter inductor state of GFM IBR \\
$x_{4d},x_{4q}$ & $d$-axis and $q$-axis component of filter capacitor state of GFM IBR \\
$x_{5d},x_{5q}$ & $d$-axis and $q$-axis component of current controller state of GFL IBR \\
$x_{pll}$ & PLL controller state of GFL IBR \\
$v_d,v_q$ & $d$-axis and $q$-axis component of voltage at point of interconnection (POI) of GFL IBR \\
$v_{d,m},v_{q,m}$ & $d$-axis and $q$-axis component of voltage following measurement delay at POI of GFL IBR \\
$u_d,u_q$ & $d$-axis and $q$-axis component of control input \\
$x_v$ & Voltage controller state of GFL IBR \\
$x_f$ & Voltage feedback delay of GFL IBR \\
$T_{mr}$ & Mechanical input at $r$th mass of multi-mass system \\
$u_g$ & Input torque at SG governor system \\
$v_{pnz}$ & Positive, negative, and zero sequence component of voltage at POI of SG \\
$i_{tpnz}$ & Positive, negative, and zero sequence component of SG terminal current\\
$v_r = [-v_f~0~0~0]^T$ & Voltage across field and damper windings of SG rotor\\
$i_r = [i_f~i_h~i_g~i_k]^T$ & Current through field and damper windings of SG rotor\\
$L_{ss}$ & Stator self-inductance matrix\\
$L_{sr},L_{rs}$ & Mutual inductance matrix between stator and rotor\\
$L_{rr}$ & Rotor self-inductance matrix \\
$T$ & $abc$ to $pnz$ transformation matrix \\
$\psi_D,\psi_Q$ & $D$-axis and $Q$-axis component of stator flux\\
$i_{tD},i_{tQ}$ & $D$-axis and $Q$-axis component of SG terminal current\\
$v_{ref}$ & Reference terminal voltage of SG\\
$v_f,e_{fd}$ & Excitation voltage in reciprocal and non-reciprocal per unit system\\
$x_s$ & Rotor speed measurement sensor state\\
$x_{wo}$ & Washout block state \\
$x_{comp1},x_{comp2}$ & PSS lead-lag compensator block states \\
$i_{cpnz}$ & Positive, negative, and zero sequence component of current through transmission line capacitance \\
$i_{LLpnz}$ & Positive, negative, and zero sequence component of current through load inductance \\
$i_{LRpnz}$ & Positive, negative, and zero sequence component of current through load resistance \\
$i_{DCpnz}$ & Positive, negative, and zero sequence component of current through AI DC load \\
$P_{DC},Q_{DC}$ & Active and reactive power of AI DC load\\

\end{longtable}

\newpage

\begin{table*}[!ht]
\centering
\caption{Summary of literature review}
\label{tab:summary}
\begin{tabular}{|>{\centering\arraybackslash} p{1.5cm}|>{\centering\arraybackslash} p{1.8cm}|>{\centering\arraybackslash} p{2.7cm}|>{\centering\arraybackslash} p{1.7cm}|>{\centering\arraybackslash} p{1.9cm}|>{\centering\arraybackslash} p{2.4cm}|}
\hline
\textbf{Literature} &
\textbf{System size} &
\textbf{Multi-mass turbine + torsional interaction} &
\textbf{Detailed IBR modeling} &
\textbf{IBR-induced SSO modeling} &
\textbf{Unbalanced fault + current limiting} \\
\hline

[5] &
SG: 1 Bus: 4 &
$\times$ &
$\times$ &
$\times$ &
$\times$ \\
\hline

[6] &
SG: 1 Bus: 4 &
\checkmark &
$\times$ &
$\times$ &
$\times$ \\
\hline

[7] &
IBR: 1 Bus: 4 &
$\times$ &
\checkmark &
$\times$ &
\checkmark \\
\hline

[8] &
SG: 3 IBR: 3 Bus: 9 &
$\times$ &
$\times$ &
$\times$ &
$\times$ \\
\hline

[9] &
SG: 2 Bus: 4 &
$\times$ &
$\times$ &
$\times$ &
$\times$ \\
\hline

[10] &
SG: 10 Bus: 39 &
$\times$ &
$\times$ &
$\times$ &
$\times$ \\
\hline

[11] &
SG: 2 IBR: 2 Bus: 11 &
$\times$ &
\checkmark &
\checkmark &
\checkmark \\
\hline

Proposed DP framework &
SG: 16 IBR: 2 Bus: 70 &
\checkmark &
\checkmark &
\checkmark &
\checkmark \\
\hline

\end{tabular}
\end{table*}

\small
\begin{table*}[!ht]
\centering
\caption{parameters of GFL IBR model}
\vspace{-5pt}
    \label{tab:list_of_para}
    \begin{tabular}{c {l}c c}
         \hline
         \hline
         \textbf{Symbol} & \textbf{Description} & \textbf{Value} \\
         \hline
         $f_{s}$ & Nominal frequency ($Hz$) & $60$\\
         $R$ & AC side filter resistance ($m\Omega$) &  $1.446$\\
         $L$ & AC side filter inductance ($mH$) &  $0.289$\\         
         $P_c^*$ & Reference real power ($MW$) &  $700$\\
         $k_p$ & Proportional gain of PLL controller ($rads^{-1}/pu$) & $101$ \\
         $k_i$ & Integral gain of PLL controller ($rads^{-1}/pu/s$)  & $2562$ \\
         $k_{p,v}$ & Proportional gain of outer control loop ($Mvar/pu$) & $4500$\\
         $k_{i,v}$ & Proportional gain of outer control loop ($Mvar/pu/s$) & $450$\\
         $k_{p,i}$ & Proportional gain of inner current control loop ($V/A$) & $0.373$\\
         $k_{i,i}$ & Integral gain of inner current control loop ($V/A/s$) & $1.157$\\
         $\tau_m$ & Voltage measurement delay ($ms$) & $1$\\
         $\tau_m$ & Voltage feedback delay ($ms$) & $50$\\
         \hline \vspace{-5pt}
    \end{tabular}
\end{table*}
\normalsize

\textbf{Note:} Parameters of GFM IBR model is taken from [7]. All other multi-mass parameters except damping are taken from [14] for G3, G4, and G6 and from [20, p.1040] for G2, G5, G7, G8, and G10. Damping parameters for (G3, G4, G6) are $D_{GE} = 0.002$, $\bar{D}_{LPB} = 1.6$, and $\bar{D}_I = 0.1$. Similarly, damping parameters for other generators are $D_{GE} = 0.002$, $\bar{D}_{LPA} = 1.3$, $\bar{D}_{LPB} = 1.6$, and $\bar{D}_I = 0.1$. Other damping parameters are set to zero. All the values are in p.u.

        

\bibliography{Mybib}